\documentclass[11pt, a4paper]{article}
\usepackage{jheppub, tensind}

\tensordelimiter{?}
\newcommand{\N}{{\mathcal N}}
\newcommand{\eol}{\notag \\}
\newcommand{\lc}{{\vert}}

\newcommand{\sym}[1]{\stackrel{\scriptstyle#1}{\mbox{\tiny $\smile$}}}

\newcommand{\sdet}{{\textrm{sdet}}}

\newcommand{\cF}{{\mathcal F}}
\newcommand{\cH}{{\mathcal H}}
\newcommand{\cM}{{\mathcal M}}
\newcommand{\cR}{{\mathcal R}}

\newcommand{\chM}{{\mathfrak M}}

\newcommand{\cha}{{\mathfrak a}}

\newcommand{\chm}{{\mathfrak m}}

\newcommand{\cE}{{\mathcal E}}

\newcommand{\bphi}{{\bar\phi}}
\newcommand{\bpsi}{{\bar\psi}}
\newcommand{\cW}{{\mathcal W}}

\newcommand{\chz}{{\mathfrak z}}

\newcommand{\balpha}{{\boldsymbol\alpha}}
\newcommand{\bbeta}{{\boldsymbol\beta}}
\newcommand{\bgamma}{{\boldsymbol\gamma}}
\newcommand{\bdelta}{{\boldsymbol\delta}}
\newcommand{\dbalpha}{{\boldsymbol\dalpha}}
\newcommand{\dbbeta}{{\boldsymbol\dbeta}}
\newcommand{\dbgamma}{{\boldsymbol\dgamma}}
\newcommand{\dbdelta}{{\boldsymbol\ddelta}}
\newcommand{\bmu}{{\boldsymbol\mu}}
\newcommand{\dbmu}{{\boldsymbol\dmu}}
\newcommand{\bnu}{{\boldsymbol\nu}}
\newcommand{\dbnu}{{\boldsymbol\dnu}}

\newcommand{\beq}{\begin{equation}}
\newcommand{\eeq}{\end{equation}}
\newcommand{\ul}{\underline}

\newcommand{\eps}{{\epsilon}}

\newcommand{\hc}{\mathrm{h.c.}}
\newcommand{\Lag}{\mathcal L}

\newcommand{\CD}{\mathcal{D}}
\newcommand{\BCD}{\mathcal{\bar D}}

\newcommand{\poin}[1]{\hat{#1}}

\newcommand{\bsigma}{\bar{\sigma}}
\newcommand{\dalpha}{{\dot{\alpha}}}
\newcommand{\dbeta}{{\dot{\beta}}}
\newcommand{\dgamma}{{\dot{\gamma}}}
\newcommand{\ddelta}{{\dot{\delta}}}
\newcommand{\dmu}{{\dot{\mu}}}
\newcommand{\dnu}{{\dot{\nu}}}
\newcommand{\dphi}{{\dot{\phi}}}
\newcommand{\btheta}{{\bar\theta}}

\newcommand{\chE}{\mathcal E}

\newcommand{\1}{{\ul 1}}
\newcommand{\2}{{\ul 2}}
\newcommand{\A}{{\mathbb A}}
\newcommand{\D}{{\mathbb D}}

\title{\boldmath $\N=2$ Conformal Superspace in Four Dimensions}
\author{Daniel Butter}
\affiliation{School of Physics M013, The University of Western Australia \\
35 Stirling Highway, Crawley, W.A. 6009, Australia}
\emailAdd{dbutter@cyllene.uwa.edu.au}

\abstract{We develop the geometry of four dimensional $\N=2$ superspace
where the entire conformal algebra of $\rm SU(2,2\vert 2)$ is realized linearly
in the structure group rather than just the $\rm SL(2,\mathbb C) \times \rm U(2)_R$
subgroup of Lorentz and $R$-symmetries, extending to $\N=2$ our prior result
for $\N=1$ superspace. This formulation explicitly lifts to superspace the existing
methods of the $\N=2$ superconformal tensor calculus;
at the same time the geometry, when degauged to $\rm SL(2,\mathbb C) \times \rm U(2)_R$,
reproduces the existing formulation of $\N=2$ conformal supergravity constructed by
Howe.}

\begin{document}
\maketitle
\section{Introduction}

It has long been apparent that superconformal techniques have
a key role to play in constructing supergravity
theories.\footnote{See for example the comprehensive review article by Fradkin and
Tseytlin \cite{csg}.}
For $\N=1$ theories the role is quite well understood: any action
involving supergravity coupled to matter can be described (often
more easily!) by conformal supergravity coupled to some compensating field
plus matter. Even in the absence of matter, quantizing supergravity
is most easily done via the introduction of compensator fields, a topic
which by now is textbook material \cite{superspace, BK}.
In the presence of matter, the compensator fields and conformal techniques
are even more important \cite{butter2}.

The same can be said of $\N=2$ supergravity theories, although they are
not nearly so well understood as their $\N=1$ cousins. 
A central tool in their analysis is the so-called superconformal
tensor calculus, which for brevity's sake we will occasionally refer
to as STC. This formalism, which allows the construction of component
actions which respect superconformal invariance, was given for
$\N=2$ supergravity by de Wit, van Holten, and Van Proeyen \cite{sct_rules},
who along with various collaborators throughout the 1980s made
extensive use of this technique \cite{sct_chiral,
sct_structure, sct_imptensor, sct_gaugematter, sct_lagrangians}; for example,
the various $\N=2$ supergravity theories can be easily
understood in the STC framework by coupling conformal supergravity
to different compensator fields \cite{sct_imptensor}.
This method continues to be used to
this day \cite{LopesCardoso:2000qm,deWit:2006gn,deWit:2010za}.

Superspace techniques also exist for $\N=2$ supergravity
in several flavors.\footnote{See for example 
\cite{BS:minimal_multiplet} and \cite{n2_superspace}, as well as
\cite{Gates:1980ky} for an an early discussion of superconformal aspects of $\N=2$
supergravity in superspace.}
The approaches most closely related to the one we will take 
are that of Grimm \cite{Grimm}, who chose $\rm SL(2,\mathbb C) \times \rm SU(2)_R$
for the structure group, and Howe \cite{Howe}, who extended this work to general $\N$ with the
structure group $\rm SL(2,\mathbb C) \times \rm U(\N)_R$.
These superspace formulations have recently been elaborated
upon and used to describe supergravity in projective
superspace \cite{KLRT_M_Grimm, KLRT_M_Howe}.\footnote{It was
shown in \cite{KLRT_M_Howe} that Grimm's formulation is a gauge-fixed
version of Howe's.}
Although such superspace techniques allow a certain super-Weyl
transformation, the relation to the \emph{manifestly} superconformal method
of de Wit et al. has remained opaque, especially since the
super-Weyl transformation is manifested inhomogeneously on the
torsion superfields.

In this paper we will clarify this relation by constructing in
superspace the \emph{manifestly} superconformal geometry corresponding
to $\N=2$ conformal supergravity. That is, we will present 
a superspace, which we call $\N=2$ conformal superspace,
where the structure group is $\rm SU(2,2\vert 2)$ and whose
component form reduces to $\N=2$ superconformal tensor calculus.
Since Howe's formulation is well known to correspond to conformal
supergravity with an unconstrained dilatation parameter \cite{Howe, KLRT_M_Howe},
it must (and, we will show, does) correspond to a certain
gauge-fixed version of the theory we construct. These results
extend to $\N=2$ our previous work \cite{Butter:2009cp}
where we constructed $\N=1$ conformal superspace.

Prior experience with superspace might hint that a larger structure
group would necessarily yield a more complicated theory.
However, as we showed in \cite{Butter:2009cp},
the Bianchi identities and curvature structure of the $\N=1$ theory
were actually \emph{simpler} and superficially resembled super
Yang-Mills.\footnote{More accurately, the constraints were
the usual integrability conditions for the existence of covariantly
chiral superfields.}
The same holds in $\N=2$, where we will show that the
constraint structure of $\N=2$ conformal supergravity may be
described by covariant derivatives with a simple algebra,
\begin{gather*}
\{\nabla_\alpha{}^i, \nabla_\beta{}^j\} = -2 \, \eps^{ij} \eps_{\alpha \beta} \bar \cW, \qquad
\{\bar\nabla^\dalpha{}_i, \bar\nabla^\dbeta{}_j\} = +2 \, \eps_{ij} \eps^{\dalpha \dbeta} \cW \\
\{\nabla_\alpha{}^i, \bar\nabla_{\dbeta j}\} = -2i \delta^i_j \nabla_{\alpha \dbeta}
\end{gather*}
where $\cW$ is an $\N=2$ field strength valued in the superconformal
algebra and obeying certain constraints. As with our $\N=1$ construction,
the resemblance to super Yang-Mills significantly simplifies the
algebra of covariant derivatives and finding solutions to the Bianchi identities.

This paper is organized as follows. In section \ref{Section2}, we construct a superspace
geometry with the superconformal algebra as its structure group and impose
a set of constraints on the curvatures analogous to the constraints imposed
in the $\N=1$ theory. In section \ref{Section3}, we describe in detail how the
superconformal tensor calculus
\cite{sct_rules, sct_chiral, sct_imptensor, sct_structure, sct_gaugematter, sct_lagrangians}
corresponds to the component version
of the superspace theory. In section \ref{Section4}, we demonstrate how the
superconformal structure may be ``degauged'' to reproduce the $\N=2$ conformal
supergravity structure found by Howe in superspace \cite{Howe}.
In the conclusion, we briefly speculate about possible extensions of this formalism.

\section{$\N=2$ conformal superspace}\label{Section2}
In this section we present our construction of $\N=2$ conformal superspace,
the superspace structure which arises from the gauging of the $\N=2$
superconformal algebra. In all of the salient details, the construction
parallels our previous work on $\N=1$ conformal superspace \cite{Butter:2009cp},
but we will attempt to present the material in a self-contained way. We begin with
defining the $\N=2$ superconformal algebra to fix our notation. Next, we summarize
the gauging procedure. Then we explain the curvature constraints which correspond
to $\N=2$ conformal supergravity. We finish with a summary of the $\N=2$
supersymmetric action principles.

\subsection{The global $\N=2$ superconformal algebra}
The global $\N=2$ superconformal algebra can be constructed from the $\N=2$
super-Poincar\'e algebra,
\begin{gather}
\{Q_\alpha{}^i, \bar Q_{\dalpha j} \} = -2i \,\delta^i_j \,\sigma^a_{\alpha \dalpha}  \,P_a \eol{}
[M_{ab}, M_{cd}] = \eta_{bc} M_{ad} - \eta_{ac} M_{bd} - \eta_{bd} M_{ac} + \eta_{ad} M_{bc}\eol{}
[M_{ab}, P_c] = P_a \eta_{bc} - P_b \eta_{ac} \eol{}
[M_{ab}, Q_\gamma{}^i] = {(\sigma_{ab})_\gamma}^{\beta} Q_\beta{}^i, \quad
[M_{ab}, \bar Q^\dgamma{}_i] = {(\bsigma_{ab})^\dgamma}_{\dbeta} \bar Q^\dbeta{}_i
\end{gather}
with all other commutators vanishing. The supersymmetry generator $Q_\alpha{}^i$
lies in the chiral spinor representation of the Lorentz group\footnote{We use
the same notation for representations of the Lorentz group as in
\cite{Butter:2009cp} and \cite{bgg}. For representations of the $\rm SU(2)_R$
algebra, we follow the conventions of \cite{KLRT_M_Grimm, KLRT_M_Howe}.} and
in the isospinor representation of $\rm SU(2)_R$. As usual, conjugate representations
of the $\rm SU(2)_R$ group are related by raising and lowering with the antisymmetric
tensor $\eps_{ij}$. Our conventions are
\begin{align}
\eps^{\1\2} = \eps_{\2\1} = 1, \qquad Q_{\alpha\, i} = \eps_{ij} Q_\alpha{}^j, \qquad
     \bar Q_\dalpha{}^i =\eps^{ij} \bar Q_{\dalpha j}.
\end{align}

The bosonic part of this algebra can be extended to include the conformal algebra.
This requires the introduction of the dilatation operator $\D$ and the
special conformal operator $K_a$. Consistency requires the further introduction
of three new operators: the fermionic special conformal operator $S_{\alpha i}$
(and its conjugate $\bar S^{\dalpha i}$), as well as the $\rm U(1)_R$ chiral rotation
operator $\A$ and the $\rm SU(2)_R$ isospin operator $I^i{}_j = I_j{}^i$ which together span the
R-symmetry group $\rm U(2)_R$.

The special conformal and superconformal generators $K$ and $S$ have obvious
Lorentz and $\rm SU(2)_R$ transformation properties, while their dilatation and $\rm U(1)_R$
properties are opposite those of $P$ and $Q$:
\begin{gather}
[\D,P_a] = P_a, \quad [\D, Q_\alpha{}^i] = \frac{1}{2} Q_\alpha{}^i, \quad
     [\D, \bar Q^\dalpha{}_i] = \frac{1}{2} \bar Q^\dalpha{}_i \eol{} 
[\D,K_a] = -K_a, \quad [\D, S_{\alpha i}] = -\frac{1}{2} S_{\alpha i}, \quad
     [\D, \bar S^{\dalpha i}] = -\frac{1}{2} \bar S^{\dalpha i} \eol{}
[\A, Q_\alpha{}^i] = -i Q_\alpha{}^i,\quad [\A, \bar Q^\dalpha{}_i] = +i \bar Q^\dalpha{}_i \eol{} 
[\A, S_{\alpha i}] = +i S_{\alpha i}, \quad [\A, \bar S^{\dalpha i}] = -i \bar S^{\dalpha i} \eol{}
[I^i{}_j, Q_\alpha{}^k] = \delta^k_j Q_\alpha{}^i - \frac{1}{2} \delta^i_j Q_\alpha{}^k, \quad
     [I^i{}_j, \bar Q^\dalpha{}_k] = -\delta_k^i \bar Q^\dalpha{}_j + \frac{1}{2} \delta_i^j \bar Q^\dalpha{}_k \eol{}
[I^i{}_j, S_{\alpha k}] = -\delta^i_k S_{\alpha j} + \frac{1}{2} \delta^i_j S_{\alpha k}, \quad
     [I^i{}_j, \bar S^{\dalpha k}] = \delta_j^k \bar S^{\dalpha i} - \frac{1}{2} \delta_j^i \bar S^{\dalpha k} \eol{}
[M_{ab}, K_c] = K_a \eta_{bc} - K_b \eta_{ac} \eol{}
[M_{ab}, S_{\gamma i}] = {(\sigma_{ab})_\gamma}^{\beta} S_{\beta i}.
\end{gather}
The special conformal generators have an algebra among each other that is
similar to the supersymmetry algebra:
\begin{align}
\{S_{\alpha i}, \bar S_\dalpha{}^j \} = +2i \,\delta_i^j \,\sigma^a_{\alpha \dalpha} \,K_a.
\end{align}
Finally, the commutators of the special conformal generators with the translation
and supersymmetry generators are
\begin{gather}
[K_a, P_b] = 2 \eta_{ab} \D - 2 M_{ab} \eol{}
\{S_{\alpha i}, Q_\beta{}^j\} = 2 \delta_i^j \epsilon_{\alpha \beta} \D
     - 2 \delta_i^j M_{\alpha \beta}
     - i \delta_i^j \epsilon_{\alpha \beta} \A
     - 4 \eps_{\alpha \beta} I_i{}^j \eol{}
\{\bar S^{\dalpha i}, \bar Q^\dbeta{}_j \} = 2 \delta^i_j \epsilon^{\dalpha \dbeta} \D
     - 2 \delta^i_j M^{\dalpha \dbeta}
     + i \delta^i_j \epsilon^{\dalpha \dbeta} \A 
     + 4 \eps^{\dalpha \dbeta} I^i{}_j \eol{}
[K_a, Q_\alpha{}^i] = i \sigma_{a \alpha \dbeta} \,\bar S^{\dbeta i}, \;\;\;
[K_a, \bar Q^\dalpha{}_i] = i \bsigma_a^{\dalpha \beta} \,S_{\beta i} \eol{}
[S_{\alpha i}, P_a] = i \sigma_{a \alpha \dbeta} \,\bar Q^{\dbeta}{}_i,\;\;\;
[\bar S^{\dalpha i}, P_a] = i \bsigma_a^{\dalpha \beta} \,Q_{\beta}{}^i.
\end{gather}
All other commutators vanish.

We have made use of the convenient shorthand
\begin{align}
M_{\alpha \beta} &\equiv -\eps_{\beta \gamma} (\sigma^{ba})_{\alpha}{}^\gamma M_{ab}
     = -(\sigma^{ba})_{\alpha \beta} M_{ab} \eol
M^{\dalpha \dbeta} &\equiv -\eps^{\dbeta \dgamma} (\bsigma^{ba})^{\dalpha}{}_\dgamma M_{ab}
     = -(\bsigma^{ba})^{\dalpha \dbeta} M_{ab}.
\end{align}
These are projections of the Lorentz generator; $M_{\alpha \beta}$ rotates
undotted spinors while $M^{\dalpha \dbeta}$ rotates dotted spinors:
\begin{align*}
[M_{\alpha \beta}, Q_{\gamma}{}^i] &= -Q_{\alpha}{}^i \eps_{\beta \gamma} - Q_{\beta}{}^i \eps_{\alpha \gamma} \\
[M^{\dalpha \dbeta}, \bar Q^{\dgamma}{}_i] &= -\bar Q^{\dalpha}{}_i \eps^{\dbeta \dgamma} - \bar Q^{\dbeta}{}_i \eps^{\dalpha \dgamma} \\
[M_{\alpha \beta}, P_{\gamma \dgamma}] &= -P_{\alpha \dgamma} \eps_{\beta \gamma}
     - P_{\beta \dgamma} \eps_{\alpha \gamma}, \qquad P_{\gamma \dgamma} \equiv \sigma^c_{\gamma \dgamma} P_c.
\end{align*}
Note that for this definition of $M$,
\begin{align}
\frac{1}{2} \lambda^{ab} M_{ba} = \frac{1}{2} \lambda^{\alpha \beta} M_{\beta \alpha}
     + \frac{1}{2} \lambda_{\dalpha \dbeta} M^{\dbeta \dalpha}
\end{align}
using the decomposition rule\footnote{This implies that $\lambda_{\alpha \beta}$ and $\lambda_{\dalpha \dbeta}$ are anti-conjugate to each other using the rule for complex conjugation
given in \cite{BK}.}
\begin{align}\label{eq_LorentzDecomposition}
\lambda_{\alpha \dalpha\, \beta \dbeta} 
     = 2 \eps_{\dalpha \dbeta} \lambda_{\alpha \beta}
     - 2 \eps_{\alpha \beta} \lambda_{\dalpha \dbeta}.
\end{align}

It is common for spinor indices to come paired with isospin indices, so we
introduce boldface notation $\balpha$ to encompass both. For example, we can
denote the supersymmetry generators by
\[
Q_\balpha = Q_\alpha{}^i, \quad
\bar Q^\dbalpha = \bar Q^\dalpha{}_i.
\]
Similarly, we write $\xi Q$ and $\bar \xi \bar Q$ as shorthand for
\[
\xi^\balpha Q_\balpha = \xi^\alpha{}_i Q_\alpha{}^i, \quad
\bar\xi_\dbalpha \bar Q^\dbalpha = \bar\xi_\dalpha{}^i \bar Q^\dalpha{}_i.
\]
It is convenient to introduce the symbols
\begin{gather}
C_{\balpha \bbeta} = \eps_{\alpha \beta} \eps^{ij}, \quad
C^{\balpha \bbeta} = \eps^{\alpha \beta} \eps_{ij} \eol
C_{\dbalpha \dbbeta} = \eps_{\dalpha \dbeta} \eps^{ij}, \quad
C^{\dbalpha \dbbeta} = \eps^{\dalpha \dbeta} \eps_{ij}
\end{gather}
for raising and lowering both sets of indices simultaneously;
for example,
\[
\xi_\balpha = C_{\balpha \bbeta} \xi^{\bbeta} = \xi_\alpha{}^i, \quad
\bar\xi^\dbalpha = C^{\dbalpha \dbbeta} \bar\xi_{\dbbeta} = \bar\xi^\dalpha{}_i.
\]
Similarly we may introduce $\delta_\balpha{}^\bbeta$ with the definition
\[
\delta_\balpha{}^\bbeta = \delta_\alpha{}^\beta \delta^i_j, \quad
\delta^\dbalpha{}_\dbbeta = \delta^\dalpha{}_\dbeta \delta_i^j.
\]
It follows that
\[
C_{\balpha \bbeta} C^{\bbeta \bgamma} = \delta_\balpha{}^\bgamma.
\]

Finally, we mention one last convention. We symmetrize isospin and
Lorentz indices with an appropriate factor of $n!$; for example,
\begin{align*}
V_{(i} W_{j)} = \frac{1}{2} V_i W_j + \frac{1}{2} V_j W_i, \qquad
V_{(\alpha} W_{\beta)} = \frac{1}{2} V_\alpha W_\beta + \frac{1}{2} V_\beta W_\alpha.
\end{align*}
However, we \emph{do not} include the $n!$ when \emph{antisymmetrizing}.
For example,
\begin{align*}
V_{[a} W_{b]} = V_a W_b - V_b W_a.
\end{align*}
This is so that (among other reasons) we can write curvatures as
$F_{nm} = \partial_{[n} A_{m]}$ without additional factors of $2$.
We follow the same convention when defining graded antisymmetrization:
$F_{NM} = \partial_{[N} A_{M]} = \partial_N A_M - \partial_M A_N (-)^{nm}$.

\subsection{$\N=2$ superspace and gauging the $\N=2$ superconformal algebra}
The natural domain in which to deal with $\N=2$ supersymmetry
is the four dimensional $\N=2$ superspace $\cM^{4 \vert 8}$ parametrized
by the local coordinates
$z^M = (x^m, \theta^\bmu, \btheta_\dbmu ) = (x^m, \theta^\mu{}_i, \btheta_\dmu{}^i)$
where $m = 0, 1, 2, 3$, $\mu = 1,2$, $\dmu = 1,2$, and $i = \ul 1, \ul 2$.
To encode supergravity, the geometry of $\N=2$ superspace must be nontrivial;
that is, we must have a nontrivial vielbein and non-vanishing connections.
However, gauging the superconformal algebra is a less trivial task than
gauging super Yang-Mills or even the Lorentz algebra. The reason for
this is that the (graded) commutator of the special (super)conformal generator
$K^A = (K^a, S^{\balpha}, \bar S_\dbalpha)$ with
$P_A = (P_a, Q_\balpha, Q^\dbalpha)$ gives generators other than
$P_A$; this means that other connections will, under the action of $K$,
transform into the vielbein. Moreover, $P_A$ cannot quite be the
same as the covariant derivative since the (graded) commutator of $P_A$
with itself corresponds to a flat geometry.\footnote{We will present a point of
view of how to gauge such an algebra which seems to differ in interpretation
from other approaches reviewed for example in \cite{csg}. (We
elaborated more fully on our interpretation in \cite{Butter:2009cp}.) However,
it should be emphasized that the difference is \emph{merely} interpretation;
the final geometric structure arrived at is the same.}

Because we have a good number of generators, it will be useful to use a
compact notation. We will denote the elements $\D$, $\A$, $I^i{}_j$,
$M_{ab}$, $K_A$ by the generic notation $X_{\ul a}$. It should be
emphasized that this set closes among itself under (graded) commutation;
we denote this set $\cH$. The remaining generators are the super-translation
generators $P_A$. The algebra of all these elements may be written
\begin{subequations}\label{eq_globalC}
\begin{gather}
[X_{\ul a}, X_{\ul b}] = -f_{\ul a \ul b}{}^{\ul c} X_{\ul c} \\
[X_{\ul a}, P_B] = -f_{\ul a B}{}^{\ul c} X_{\ul c} - f_{\ul a B}{}^{C} P_{C} \\
[P_A, P_B] = -T_{AB}{}^C P_C 
\end{gather}
\end{subequations}
where the $f$'s are structure constants and $T_{AB}{}^C$ is the constant
torsion tensor. We will \emph{gauge} such an algebra by constructing an operator
$\nabla_A$, called the \emph{covariant derivative}, which plays the role
of $P_A$ in the above relations, with the caveat that the last relation
is relaxed to something more general:
\begin{align}
[\nabla_A, \nabla_B] = -T_{AB}{}^C \nabla_C - R_{AB}{}^{\ul c} X_{\ul c}
\end{align}
where $T_{AB}{}^C$ is a more general torsion tensor and $R_{AB}{}^{\ul c}$
a curvature tensor.

We begin by associating a connection one-form $H_M{}^{\ul a}$ with each
generator $X_{\ul a}$ we wish to gauge. In addition, we introduce the vielbein
one-form $E_M{}^A$ with the usual physical interpretation of equipping the
manifold $\cM^{4 \vert 8}$ with a tangent frame. Under the gauge transformations $\cH$ associated
with $X_{\ul a}$, we take
\begin{subequations}
\begin{align}\label{eq_Econn}
\delta_\cH E_M{}^A &= E_M{}^B g^{\ul c} f_{\ul c B}{}^A \\ \label{eq_Hconn}
\delta_\cH H_M{}^{\ul a} &= \partial_M g^{\ul a}
     + E_M{}^B g^{\ul c} f_{\ul c B}{}^{\ul a}
     + H_M{}^{\ul b} g^{\ul c} f_{\ul c \ul b}{}^{\ul a}
\end{align}
\end{subequations}
where $f_{\ul c B}{}^A$ and $f_{\ul c B}{}^{\ul a}$ are the structure constants
from \eqref{eq_globalC}.
 
A \emph{covariant} superfield $\Psi$ is defined by the property that
it transforms under gauge transformations $\cH$ without any derivative
on the parameter $g^{\ul a}$,
\begin{align}
\delta_\cH \Psi = g^{\ul a} X_{\ul a} \Psi.
\end{align}
The operator $X_{\ul a}$ acts on $\Psi$, transforming this local
field into some other local field. For example, if $\Psi$ is
a conformally \emph{primary} superfield, we have
\begin{align}
K_A\Psi = 0,
\end{align}
whereas if $\Psi$ is a \emph{descendant} of some other fields $\Phi_A$,
we have
\begin{align}
K_A \Psi = \Phi_A.
\end{align}
For the other generators, we normally have the usual matrix
representations
\begin{gather}
\D \Psi = \Delta \Psi, \quad
\A \Psi = i w \Psi, \quad
I^i{}_j \Psi = \mathcal J^i{}_j \Psi, \quad
M_{ab} \Psi = \mathcal S_{ab} \Psi, \quad
\end{gather}
where $\Delta$ and $w$ are real numbers, corresponding to the
conformal dimension and $\rm U(1)_R$ weights of $\Psi$, while
$\mathcal J^i{}_j$ and $\mathcal S_{ab}$ are the isospin and
Lorentz matrices associated with $\Psi$'s representation.

Because the parameter $g^{\ul a}$ is a local superfield,
$\partial_M \Psi$ does not transform covariantly. We must
introduce instead the \emph{covariant derivative}
\begin{align}
\nabla_A \Psi \equiv E_A{}^M \partial_M \Psi - E_A{}^M H_M{}^{\ul a} X_{\ul a} \Psi.
\end{align}
The superfield $\nabla_A \Psi$ is also covariant; one can show that it
transforms as
\begin{align}
\delta_\cH (\nabla_A \Psi) = g^{\ul b} \nabla_A X_{\ul b} \Psi
     - g^{\ul b} f_{\ul b A}{}^C \nabla_C \Psi
     - g^{\ul b} f_{\ul b A}{}^{\ul c} X_{\ul c} \Psi
\end{align}
without any derivatives of $g$.
If we denote $\delta_\cH (\nabla_A \Psi) = g^{\ul b} X_{\ul b} \nabla_A \Psi$,
we immediately find the operator relation
\begin{align}
[X_{\ul b}, \nabla_A] = -f_{\ul b A}{}^C \nabla_C
     - f_{\ul b A}{}^{\ul c} X_{\ul c}.
\end{align}

The \emph{curvatures} of the manifold $\cM$ are found by taking
the commutator of two covariant derivatives,
\begin{align}
[\nabla_A, \nabla_B] \Psi = -T_{AB}{}^C \nabla_C \Psi - R_{AB}{}^{\ul c} X_{\ul c} \Psi
\end{align}
where $T$ and $R$ are the torsion and $\cH$-curvature two-forms, respectively.
They are most easily written as two-forms:
\begin{subequations}
\begin{align}\label{eq_Tcurv}
T^A &\equiv d E^A - E^B \wedge H^{\ul c} \,f_{\ul c B}{}^A \\ \label{eq_Rcurv}
R^{\ul a} &\equiv d H^{\ul a} - E^B \wedge H^{\ul c} \,f_{\ul c B}{}^{\ul a}
     - \frac{1}{2} H^{\ul b} \wedge H^{\ul c} \,f_{\ul c \ul b}{}^{\ul a}.
\end{align}
\end{subequations}

Our structure can be understood then as consisting of a set of operators
$X_{\ul a}$ and $\nabla_A$ with an algebra
\begin{subequations}
\begin{align}
[X_{\ul a}, X_{\ul b}] &= -f_{\ul a \ul b}{}^{\ul c} X_{\ul c} \\
[X_{\ul a}, \nabla_B] &= -f_{\ul a B}{}^C \nabla_C - f_{\ul a B}{}^{\ul c} X_{\ul c} \\
[\nabla_A, \nabla_B] &= -T_{AB}{}^C \nabla_C - R_{AB}{}^{\ul c} X_{\ul c}.
\end{align}
\end{subequations}
For the \emph{global} superconformal algebra, we may identify $\nabla_A$
with $P_A$ provided the $\cH$-curvatures $R$ and all torsions
(except $T_\balpha{}^{\dbbeta c}$) vanish. This is simply
the geometry of flat superspace. A curved superspace corresponds
then to a deformation of the superconformal algebra by
introducing extra structure \emph{functions} $T$ and $R$ into the
algebra in the form of curvatures.
This is the sense in which we ``gauge'' the superconformal algebra.

In order for this operator structure to be consistent, the
Jacobi identities must be satisfied. We know they are satisfied
for the global algebra, so we need to check them only for the local
case. The Jacobi identites involving at most one $\nabla$ are
\begin{align*}
0 = [X_{\ul a}, [X_{\ul b}, X_{\ul c}]] + \textrm{permutations} \\
0 = [X_{\ul a}, [X_{\ul b}, \nabla_C]] + \textrm{permutations}
\end{align*}
and both are unchanged from the global case. The Jacobi identity
involving two $\nabla$'s is
\begin{align*}
0 = [X_{\ul a}, [\nabla_B, \nabla_C]] + \textrm{permutations},
\end{align*}
which implies that the curvatures transform under $\cH$ covariantly as
\begin{subequations}
\begin{align}
X_{\ul a} T_{BC}{}^{D} &=
     - T_{BC}{}^F f_{F \ul a}{}^{D}
     - f_{\ul a [B}{}^F T_{F C]}{}^{D}
     - f_{\ul a [B}{}^{\ul f} f_{\ul f C]}{}^{D} \\
X_{\ul a} R_{BC}{}^{\ul d} &=
     - T_{BC}{}^F f_{F \ul a}{}^{\ul d}
     - R_{BC}{}^{\ul f} f_{\ul {f a}}{}^{\ul d}
     - f_{\ul a [B}{}^F R_{F C]}{}^{\ul d}
     - f_{\ul a [B}{}^{\ul f} f_{\ul f C]}{}^{\ul d}.
\end{align}
\end{subequations}
One can check using the definition of the curvatures, \eqref{eq_Tcurv} and \eqref{eq_Rcurv},
along with the rules for the transformation of the connections, \eqref{eq_Econn} and
\eqref{eq_Hconn}, that the curvatures do indeed transform in this way.
Finally, the Jacobi identities involving three $\nabla$'s (i.e. the Bianchi
identities) are
\begin{align}
0 = [\nabla_A, [\nabla_B, \nabla_C]] + \textrm{permutations}.
\end{align}
We will check that these are satisfied in the next section once we postulate
the form of the curvatures.

We collect our definitions of the curvatures and connections for the
case of the superconformal algebra in Appendix \ref{app_Curvatures}.

\subsection{Conformal supergravity curvature constraints}
As with all theories in superspace, the number of degrees of freedom in
the connections far exceeds any reasonable number.
We must therefore constrain the theory. The usual way this is done is
by constraining curvatures (for super Yang-Mills theories) or torsions
(for supergravity theories) and then (1) checking the Bianchi identity and
(2) ensuring that the theory is not overconstrained.

The first of these is a difficult enough problem in superspace. Thankfully,
we already have a clue as to how to proceed. In our previous work on
$\N=1$ conformal superspace \cite{Butter:2009cp}, we found that the constraints necessary
to encode conformal supergravity were a subset of ``gauge'' constraints --
that is, constraints which superficially resemble the constraints of
super Yang-Mills. The Bianchi identity was then very easy to check since
its form was identical to the Bianchi identity of super Yang-Mills.
It turns out that the same is true of $\N=2$ conformal supergravity in
superspace. We begin then by postulating the following form for the spinor
derivative curvatures:
\begin{subequations}\label{eq_gaugeconstraints}
\begin{align}
\{\nabla_\alpha{}^i, \nabla_\beta{}^j\} &= -2 \, \eps^{ij} \eps_{\alpha \beta} \bar \cW \\
\{\bar\nabla^\dalpha{}_i, \bar\nabla^\dbeta{}_j\} &= +2 \, \eps_{ij} \eps^{\dalpha \dbeta} \cW \\
\{\nabla_\alpha{}^i, \bar\nabla_{\dbeta j}\} &= -2i \delta^i_j \nabla_{\alpha \dbeta}
\end{align}
\end{subequations}
where $\cW$ and $\bar \cW$ are valued in the superconformal algebra,
\begin{align}
\cW = \cW(P)^A \nabla_A + \frac{1}{2} \cW(M)^{ba} M_{ab}
     + \cW(\D) \D + \cW(\A) \A + \cW(I)^j{}_i I^i{}_j
     + \cW(K)^A K_A.
\end{align}
We will soon impose some of these $\cW$ to be zero, but for the moment we will consider
this more general case. As in super Yang-Mills, the operator $\cW$ must obey
two conditions in order for the Bianchi identities to be satisfied.
First, $\cW$ must be a chiral operator\footnote{It is important to note that
this condition does \emph{not} mean that the components of $\cW$ are all chiral
superfields.}
\begin{align}
[\nabla_\alpha{}^i, \bar \cW] = [\bar\nabla^\dalpha{}_i, \cW] = 0.
\end{align}
Second, it must obey the Bianchi identity
\begin{align}
\{\nabla^{\phi i}, [\nabla_{\phi j}, \cW]\} = \{\bar\nabla_{\dphi j}, [\bar\nabla^{\dphi i}, \bar \cW]\}.
\end{align}
In order for the Jacobi identities to be satisfied, $\cW$ must obey one
nontrivial constraint: it must be conformally primary
\begin{align}
[S_{\beta i}, \cW] = [S^{\dbeta j}, \cW] = 0 \implies [K_c, \cW] = 0.
\end{align}

It is straightforward to derive the form of all other curvatures in terms of the
operator $\cW$. For the dimension-3/2 curvatures, we have
\begin{subequations}
\begin{align}
[\nabla_\beta{}^i, \nabla_{\alpha \dalpha}] &= -2 \eps_{\beta \alpha} \bar\cW_{\dalpha}{}^i \\
[\bar\nabla_{\dbeta i}, \nabla_{\alpha \dalpha}] &= -2 \eps_{\beta \alpha} \cW_{\alpha i}
\end{align}
\end{subequations}
where
\begin{subequations}
\begin{align}
\cW_{\alpha i} &\equiv -\frac{i}{2} \eps_{ij} [\nabla_{\alpha}{}^j, \cW] \\
\cW_{\dalpha}{}^i &\equiv -\frac{i}{2} \eps^{ij} [\bar\nabla_{\dalpha j}, \cW].
\end{align}
\end{subequations}
The dimension-2 curvatures are
\begin{align}
[\nabla_{\beta \dbeta}, \nabla_{\alpha \dalpha}] &= -\cF_{\beta \dbeta \,\alpha \dalpha}
     = -2 \eps_{\dbeta \dalpha} \cF_{\sym{\beta \alpha}}
     + 2 \eps_{\beta \alpha} \cF_{\sym{\dbeta \dalpha}},
\end{align}
where
\begin{subequations}
\begin{align}
\cF_{\sym{\beta \alpha}} &= \frac{1}{8} \{\nabla_{(\beta}{}^k,[\nabla_{\alpha) k}, \cW]\} \\
\cF_{\sym{\dbeta \dalpha}} &= \frac{1}{8} \{\bar\nabla_{(\dbeta}{}^k,[\bar\nabla_{\dalpha) k}, \bar \cW]\}.
\end{align}
\end{subequations}
Now we must specify the precise form of $\cW$ and $\bar \cW$.
In the $\N=1$ case, the spinor superfield $\cW_\alpha$ was constrained so that
only the Lorentz and special conformal components were nonvanishing; moreover, they
were both given in terms of a single chiral superfield $W_{\alpha \beta \gamma}$,
containing the field strengths of $\N=1$ conformal supergravity. It turns out that
an analogous structure may be imposed here. We take the scalar superfield $\cW$ to be
defined in terms of the single chiral superfield $W_{\alpha \beta}$ and its derivatives,
\begin{subequations}
\begin{align}
\cW = \frac{1}{2} W^{\alpha \beta} M_{\beta \alpha} + \cW(S)^{\beta j} S_{\beta j}
     + \cW(K)^b K_b \\
\bar \cW = \frac{1}{2} \bar W_{\dalpha \dbeta} M^{\dbeta \dalpha}
     + \bar \cW(S)_{\dbeta j} \bar S^{\beta j}
     + \bar \cW(K)^b K_b
\end{align}
\end{subequations}
where
\begin{subequations}
\begin{gather}
\cW(S)_\alpha{}^i = \frac{1}{4} \nabla^{\beta i} W_{\beta \alpha}, \;\;\;
\bar \cW(S)_i{}^\dalpha = \frac{1}{4} \nabla_{\dbeta i} \bar W^{\dbeta \dalpha} \\
\cW(K)^{\dalpha \alpha} = -\frac{1}{2} \nabla^{\dalpha \beta} W_\beta{}^\alpha, \;\;\;
\bar \cW(K)_{\alpha \dalpha} = -\frac{1}{2} \nabla_{\alpha \dbeta} \bar W^\dbeta{}_\dalpha.
\end{gather}
\end{subequations}
It is straightforward to check that $\cW$ obeys all the necessary constraints
provided the superfield $W_{\alpha \beta}$ is chiral, primary, and constrained to obey
the Bianchi identity
\begin{align}\label{eq_WBianchi}
\nabla^{\alpha \beta} W_{\beta \alpha} \equiv \nabla^{\alpha k} \nabla^{\beta}_k W_{\beta \alpha}
     = \nabla^{\dalpha k} \nabla^{\dbeta}_k \bar W_{\dbeta \dalpha}.
     \equiv \bar\nabla^{\dalpha \dbeta} \bar W_{\dbeta \dalpha}
\end{align}

To show that our structure is indeed $\N=2$ conformal supergravity can be done in
two ways. The first is to show that when reduced to components, the curvatures and constraints
on the curvatures are identical to those imposed in the $\N=2$ superconformal
tensor calculus, which is well known to describe conformal supergravity in
components \cite{sct_rules, sct_chiral, sct_structure}.
The second is to show that when the structure group is degauged to $\rm SL(2,\mathbb C) \times \rm U(2)_R$,
the theory reduces to Howe's formulation of $\N=2$ conformal supergravity \cite{Howe}.
We will show both, (re)constructing the superconformal tensor calculus in
section \ref{Section3} and Howe's formulation in section \ref{Section4}.

We can now give the form of the remaining curvatures.
The dimension-3/2 curvatures are given by the spinor superfield operators
\begin{subequations}
\begin{align}
\cW_{\alpha j} &=
     \frac{i}{2} W_\alpha{}^\gamma \nabla_{\gamma j}
     - \frac{i}{4} \nabla^{\phi}{}_j W_{\phi \alpha} \D
     - \frac{1}{8} \nabla^{\phi}{}_j W_{\phi \alpha} \A
     + \frac{i}{2} \nabla^{\phi k} W_{\phi \alpha} I_{kj}
     - \frac{i}{4} \nabla^\beta{}_j W^\gamma{}_\alpha M_{\beta \gamma}
     \eol & \quad
     - \frac{i}{8} \nabla_{\alpha j} \nabla^{\phi k} W_\phi{}^\gamma S_{\gamma k}
     + \frac{1}{4} \nabla_\dbeta{}^\phi W_{\phi \alpha} \bar S^{\dbeta}{}_j
     - \frac{i}{8} \nabla_{\alpha j} \nabla_\dbeta{}^\phi W_{\phi \beta} K^{\dbeta \beta} \\
\cW^{\dalpha j} &=
     \frac{i}{2} \bar W^\dalpha{}_\dgamma \bar \nabla^{\dgamma j}
     - \frac{i}{4} \bar \nabla_{\dphi}{}^j \bar W^{\dphi \dalpha} \D
     + \frac{1}{8} \bar \nabla_{\dphi}{}^j \bar W^{\dphi \dalpha} \A
     - \frac{i}{2} \bar \nabla_{\dphi k} \bar W^{\dphi \dalpha} I^{kj}
     - \frac{i}{4} \bar\nabla_\dbeta{}^j \bar W_\dgamma{}^\dalpha M^{\dbeta \dgamma}
     \eol & \quad
     - \frac{i}{8} \bar \nabla^{\dalpha j} \nabla_{\dphi k} \bar W^\dphi{}_\dgamma \bar S^{\dgamma k}
     + \frac{1}{4} \nabla^\beta{}_\dphi \bar W^{\dphi \dalpha} S_{\beta}{}^j
     - \frac{i}{8} \bar \nabla^{\dalpha j} \nabla^\beta{}_\dphi \bar W^{\dphi \dbeta} K_{\beta \dbeta}.
\end{align}
\end{subequations}

The dimension-2 curvatures are most easily given by specifying
the components $\cF_{\sym{\beta \alpha}}$ and
$\cF_{\sym{\dbeta \dalpha}}$ of the curvature tensor
$\cF_{\beta \dbeta\, \alpha \dalpha}$ given by
\begin{align}
\cF_{ba} = T_{ba}{}^C \nabla_C + \frac{1}{2} R_{ba}{}^{dc} M_{cd}
     + R(I)_{ba}{}^i{}_j I^j{}_i
     + R(\D)_{ba} \D
     + R(\A)_{ba} \A
     + R(K)_{ba}{}^C K_C.
\end{align}

The torsion is
\begin{subequations}
\begin{align}
T_{\sym{\beta \alpha}}{}^C \nabla_C &=
     + \frac{1}{4} \nabla^\gamma{}_j W_{\beta \alpha} \nabla_\gamma{}^j \\
T_{\sym{\dbeta \dalpha}}{}^C \nabla_C &=
     - \frac{1}{4} \bar \nabla_\dgamma{}^j \bar W_{\dbeta \dalpha} \bar \nabla^\dgamma{}_j.
\end{align}
\end{subequations}
The Lorentz curvature is
\begin{subequations}
\begin{align}
\frac{1}{2} R_{\sym{\beta \alpha}}{}^{cd} M_{dc}
     &= \frac{1}{16} \nabla^{\gamma \delta} W_{\beta \alpha} M_{\delta \gamma}
     - \frac{1}{16} \nabla_{\phi\gamma} W^{\phi \gamma} M_{\alpha \beta}
     - \frac{1}{4} W_{\alpha \beta} \bar W_{\dgamma \ddelta} M^{\ddelta \dgamma} \\
\frac{1}{2} R_{\sym{\dbeta \dalpha}}{}^{cd} M_{dc}
     &= \frac{1}{16} \bar\nabla_{\dgamma \ddelta} \bar W_{\dbeta \dalpha} M^{\ddelta \dgamma}
     - \frac{1}{16} \nabla^{\dgamma \dphi} \bar W_{\dphi \dgamma} M_{\dalpha \dbeta}
     - \frac{1}{4} \bar W_{\dalpha \dbeta} W^{\gamma \delta} M_{\delta \gamma} .
\end{align}
\end{subequations}
The isospin curvature is
\begin{subequations}
\begin{align}
R(I)_{\sym{\beta \alpha}}{}^{i j}&= -\frac{1}{8} \nabla^{ij} W_{\beta \alpha} \\
R(I)_{\sym{\dbeta \dalpha}}{}^{ij} &= -\frac{1}{8} \bar\nabla^{ij} \bar W_{\dbeta \dalpha}.
\end{align}
\end{subequations}
The dilatation curvatures are
\begin{subequations}
\begin{align}
R(\D)_{\sym{\beta \alpha}} &=
     \frac{1}{16} \nabla_{\beta}{}^\gamma W_{\gamma \alpha}
     + \frac{1}{16} \nabla_{\alpha}{}^\gamma W_{\gamma \beta} \\
R(\D)_{\sym{\dbeta \dalpha}} &=
     \frac{1}{16} \bar\nabla_{\dbeta \dgamma} \bar W^{\dgamma}{}_\dalpha
     + \frac{1}{16} \bar\nabla_{\dalpha \dgamma} \bar W^{\dgamma}{}_\dbeta
\end{align}
\end{subequations}
and the $\rm U(1)_R$ curvatures are
\begin{subequations}
\begin{align}
R(\A)_{\sym{\beta \alpha}} &=
     -\frac{i}{32} \nabla_{\beta}{}^\gamma W_{\gamma \alpha}
     - \frac{i}{32} \nabla_{\alpha}{}^\gamma W_{\gamma \beta} \\
R(\A)_{\sym{\dbeta \dalpha}} &=
     + \frac{i}{32} \bar\nabla_{\dbeta \dgamma} \bar W^{\dgamma}{}_\dalpha
     + \frac{i}{32} \bar\nabla^{\dalpha \dgamma} \bar W^{\dgamma}{}_\dbeta.
\end{align}
\end{subequations}
They are related by $R(\D)_{\sym{\beta \alpha}} = 2i R(\A)_{\sym{\beta \alpha}}$
and $R(\D)_{\sym{\dbeta \dalpha}} = -2i R(\A)_{\sym{\dbeta \dalpha}}$.
Finally, the special conformal curvatures are
\begin{subequations}\label{eq_Kcurv}
\begin{align}
R(K)_{\sym{\beta \alpha}}{}^C K_C &=
     \frac{1}{32} \nabla_{\beta \alpha} \nabla^{\dgamma \delta} W_{\delta}{}^\gamma K_{\gamma \dgamma}
     + \frac{1}{96} \nabla^{ij} \nabla_{\beta i} W_\alpha{}^\gamma S_{\gamma j}
     + \frac{1}{96} \nabla^{ij} \nabla_{\alpha i} W_\beta{}^\gamma S_{\gamma j}
     \eol & \quad
     - \frac{i}{16} \nabla_{\beta j} \nabla_\dgamma{}^\gamma W_{\gamma \alpha} \bar S^{\dgamma j}
     - \frac{i}{16} \nabla_{\alpha j} \nabla_\dgamma{}^\gamma W_{\gamma \beta} \bar S^{\dgamma j}
     \eol & \quad
     - \frac{1}{8} W_{\beta \alpha} \bar \nabla_{\dphi j} \bar W^\dphi{}_\dgamma \bar S^{\gamma j}
     - \frac{1}{8} W_{\beta \alpha} \nabla_{\gamma \dphi} \bar W^\dphi{}_\dgamma K^{\dgamma \gamma} \\
R(K)_{\sym{\dbeta \dalpha}}{}^C K_C &=
     \frac{1}{32} \bar \nabla_{\dbeta \dalpha} \nabla_{\gamma \ddelta} \bar W^{\ddelta}{}_\dgamma K^{\dgamma \gamma}
     - \frac{1}{96} \bar \nabla_{ij} \bar \nabla_\dbeta{}^{j} \bar W_{\dalpha\dgamma} \bar S^{\dgamma j}
     - \frac{1}{96} \bar \nabla_{ij} \bar \nabla_\dalpha{}^{j} \bar W_{\dbeta\dgamma} \bar S^{\dgamma j}
     \eol & \quad
     + \frac{i}{16} \bar \nabla_\dbeta{}^{j} \nabla^\gamma{}_\dgamma \bar W^{\dgamma}{}_\dalpha S_{\gamma j}
     + \frac{i}{16} \bar \nabla_\dalpha{}^{j} \nabla^\gamma{}_\dgamma \bar W^{\dgamma}{}_\dbeta S_{\gamma j}
     \eol & \quad
     - \frac{1}{8} \bar W_{\dbeta \dalpha} \nabla^{\phi j} W_\phi{}^\gamma S_{\gamma j}
     - \frac{1}{8} \bar W_{\dbeta \dalpha} \nabla^{\dgamma \phi} W_\phi{}^\gamma K_{\gamma \dgamma}.
\end{align}
\end{subequations}
We summarize these relations (and give an alternative form for the special
conformal curvatures) in Appendix \ref{app_Curvatures}.

\subsection{Superconformally invariant actions}
Having derived the algebra of covariant derivatives, we turn next to the
construction of superconformally invariant actions.
Within the context of pure $\N=2$ superspace (that is, without introducing
an auxiliary $\mathbb CP^1$ or $\rm S^2$ manifold as in projective \cite{KLR, GHR, LR, LR2} or
harmonic \cite{GIKOS, GIOS} superspace), there are two types
of supersymmetric actions which may be constructed. The first involves
an integral over the full superspace,
\begin{align}
S = \int d^{12} z\, E\, V,\qquad d^{12}z \equiv d^4x\, d^4\theta \, d^4\btheta.
\end{align}
Here $E = \sdet(E_M{}^A)$ is the appropriate measure ensuring
superdiffeomorphism invariance for a real scalar superfield Lagrangian $V$.
The invariance of this action requires \cite{Butter:2009cp} that $V$ transform
trivially under each of the generators of the conformal algebra,
\begin{gather*}
\D \, V = \A \,V = I^i{}_j V = M_{ab} V = K_a V = S_{\balpha} V = \bar S^{\dbalpha} V = 0.
\end{gather*}
In other words, $V$ must be a conformally primary Lorentz and $\rm U(2)_R$ scalar
with vanishing conformal dimension.

The second type of action available is the chiral action,
\begin{align}
S_c = \int d^{8} \chz\, \cE\, U + \hc ,\qquad
d^{8}\chz \equiv d^4x\, d^4\theta, \qquad
\bar\nabla^{\dbalpha} U = 0.
\end{align}
The first term on the right involves an integration over the chiral manifold
$\chM^{4\vert 4}$ parametrized by the coordinates $\chz^\chm = (x^m, \theta^\bmu)$.
We use the gothic index $\chm$ to correspond to the set $(m, \bmu)$; similarly,
we use $\cha$ for $(a, \balpha)$. The space $\chM^{4 \vert 4}$ may be understood as a submanifold
of $\cM^{4\vert 8}$, corresponding to the $\btheta=0$ slice in
the gauge where\footnote{We will soon show this gauge exists by direct construction.}
\begin{align}\label{eq_Eantichiralgauge}
E_M{}^A = \left(\begin{array}{cc}
E_\chm{}^\cha & E_{\chm \dbalpha} \\
E^{\dbmu \cha} & E^{\dbmu}{}_{\dbalpha}
\end{array}\right)
= \left(\begin{array}{cc}
\chE_\chm{}^\cha & E_{\chm \dbalpha} \\
0 & \delta^{\dbmu}{}_{\dbalpha}
\end{array}\right).
\end{align}
The chiral measure $\chE$ is given by $\sdet(\chE_\chm{}^\cha)$.

In order for the action $S_c$ to be invariant under the superconformal group, the chiral
Lagrangian $U$ must obey
\begin{gather*}
\bar\nabla^\dalpha U = 0, \quad
\D\, U = 2 U,\quad \A\, U = 4i U, \quad I^i{}_j U = M_{ab} U = 0 \\
K_a U = S_{\balpha} U = \bar S^{\dbalpha} U = 0.
\end{gather*}
That is, $U$ must be a conformally primary Lorentz and $\rm SU(2)_R$ chiral
scalar with conformal dimension 2 and $\rm U(1)_R$ weight 4. In addition, the
consistency conditions
\begin{gather*}
\{\nabla^{\dbalpha}, \nabla^{\dbbeta}\} U = 0
\end{gather*}
must be obeyed; for the curvatures we have imposed, we indeed have
\begin{align}
0 = \{\nabla^{\dalpha}{}_i, \nabla^{\dbeta}{}_j\} U &= 2 \eps_{ij} \eps^{\dalpha \dbeta}
     \left(\frac{1}{2} W^{\gamma \delta} M_{\gamma \delta}
     + W(S)^{\gamma k} S_{\gamma k}
     + W(K)^c K_c \right) U
\end{align}
satisfied trivially since $U$ is a Lorentz scalar and conformally primary.
Similarly, we have the consistency conditions
\begin{align}
0 = \{\bar S^{\dalpha i}, \bar\nabla^{\dbeta}{}_j\} U &=
     \delta^i_j (2 \eps^{\dalpha \dbeta} \D - 2 M^{\dalpha \dbeta} + i \eps^{\dalpha \dbeta} \A) U
     + 4 \eps^{\dalpha \dbeta} I^i{}_j U.
\end{align}
which are satisfied for scalar $U$ with the given dilatation and $\rm U(1)_R$ weights.
Finally, one additional torsion constraint must be satisfied,\footnote{For a proof
of this and previous results, see section A.2.5 of \cite{Butter:2009cp}.}
\begin{align}
0 = T^{\dbalpha}{}_b{}^b - T^{\dbalpha}{}_\bbeta{}^\bbeta,
\end{align}
which is obeyed trivially for our choice of torsion tensor.

\subsubsection{Converting full superspace integrals to chiral integrals}
As in $\N=1$ superspace, one may convert an integral over full superspace to
one over chiral superspace. The globally supersymmetric result is
\begin{align}
\int d^{12} z\, V = \int d^{8}\chz \, \bar D^4 V,
\qquad \bar D^4 \equiv \frac{1}{48} \bar D^{ij} \bar D_{ij}.
\end{align}
We will prove that its locally supersymmetric generalization is simply
\begin{align}\label{eq_DtoF}
\int d^{12} z\, E\, V = \int d^{8}\chz \, \chE\, \bar \nabla^4 V,
\qquad \bar \nabla^4 \equiv \frac{1}{48} \bar \nabla^{ij} \bar \nabla_{ij}.
\end{align}

The first step to evaluating the integral is to construct a certain normal
coordinate system where the covariant derivative $\bar\nabla^\dbalpha$ is
especially simple. Given the generic form
\begin{align}
\bar\nabla^\dbalpha = E^{\dbalpha}{}^M \Big(\partial_M - H_M{}^{\ul b} X_{\ul b}\Big)
\end{align}
for the covariant derivative, along with the constraints
\begin{align}
\{\bar\nabla^{\dbalpha}, \bar\nabla^{\dbbeta}\} = 2 C^{\dbalpha \dbbeta} \,\cW^{\ul c} X_{\ul c}, \qquad
[\bar\nabla^{\dbalpha}, \cW^{\ul c} X_{\ul c}] = 0,
\end{align}
one can adopt the normal coordinate gauge
\begin{gather}
E^{\dbmu A} = \delta^{\dbmu A}, \quad
H^{\dbmu}{}^{\ul a} = -C^{\dbmu \dbnu} \btheta_{\dbnu} \,\cW^{\ul a}.
\end{gather}
This gauge choice requires the fixing of $\btheta$-dependent terms in
all of the gauge degrees of freedom (including diffeomorphisms).
In particular, this gauge choice for the vielbein is identical to the gauge discussed in
\eqref{eq_Eantichiralgauge}. For the superconformal group, the covariant
derivative $\bar\nabla^\dbalpha$ then takes the especially simple form
\begin{align}
\bar\nabla^{\dbalpha} = \bar\partial^{\dbalpha}
     - \frac{1}{2} \Omega^{\dbalpha \beta \gamma} M_{\gamma \beta}
     - F^{\dbalpha}{}^b K_b
     - F^{\dbalpha \beta j} S_{\beta j}.
\end{align}
For a primary scalar field $\Psi$, it follows that
$\bar\nabla^\dbalpha \Psi = \bar\partial^{\dbalpha} \Psi$,
$\bar\nabla^\dbbeta \bar\nabla^\dbalpha \Psi = \bar\partial^{\dbbeta} \bar\partial^{\dbalpha} \Psi$,
and so on.

This gauge choice implies
\begin{align}
E = \sdet(E_M{}^A) = \sdet(\chE_\chm{}^\cha) = \chE,
\end{align}
and one may show that $E$ is independent of $\btheta$,
\begin{align}
\partial^{\dbmu} E = \nabla^{\dbmu} E
     = E\, (\nabla^{\dbmu} E_N{}^A) E_A{}^N (-)^n
     = E\, (T^{\dbmu}{}_A{}^A (-)^a + \nabla_A E^{\dbmu A}) = 0.
\end{align}
It immediately follows that in this gauge
\begin{align}
\int d^{12} z\, E\, V =
     \int d^{8} \chz\, \chE\, \frac{1}{48} \bar\partial^{ij} \bar\partial_{ij} V
\end{align}
and since in this gauge we have
$\bar\partial^{ij} \bar\partial_{ij} V = \bar\nabla^{ij} \bar\nabla_{ij} V$,
we arrive at the expression
\begin{align}
\int d^{12} z\, E\, V
     = \int d^{8} \chz\, \chE\, \frac{1}{48} \bar\nabla^{ij} \bar\nabla_{ij} V
\end{align}
\emph{in this particular gauge}. The result \eqref{eq_DtoF}, valid in any gauge,
follows from gauge invariance of both sides of the above relation;
that is, because it holds in the specific gauge we have constructed here,
it must hold in all gauges. This is a suspiciously simple-looking result;
we will give greater credibility to it in section \ref{Howe_action} where we
show that it implies the more complicated chiral projection operator familiar
from Howe's and Grimm's formulations of $\N=2$ superspace as constructed
by M\"uller \cite{Muller}.

In $\N=1$ Poincar\'e superspace, it is possible to perform this process in reverse,
converting an integral over the chiral superspace to one over the whole superspace
using the relation
\begin{align}\label{eq_FtoDP}
\int d^4 \chz\, \chE\, U = \frac{1}{2}\int d^{8} z\, \frac{E}{R} \, U.
\end{align}
This result follows from the $\N=1$ conformal superspace relation
\begin{align}\label{eq_FtoDC}
\int d^4 \chz\, \chE\, U = -4 \int d^{8} z\, E\, \frac{X}{\bar\nabla^2 X} U
\end{align}
where $X$ is a real superfield of conformal dimension two. In this expression,
$\bar\nabla^2 X$ is chiral and primary and so the right hand side is gauge invariant.
If one degauges the conformally covariant derivative to the normal Poincar\'e covariant
derivative, one must make the replacement
\[
\bar\nabla^2 X \longrightarrow (\BCD^2 - 8 R) X.
\]
If one simultaneously gauges $X$ to unity using the dilatational gauge symmetry,
the result \eqref{eq_FtoDP} follows.

It is natural to ask whether an analogous relation may be constructed for $\N=2$
superspace. Although we haven't yet discussed how the degauging procedure
works, it is easy to see that the answer must be no. The analogue of \eqref{eq_FtoDC} is 
\begin{align}\label{eq_FtoD}
\int d^{8} \chz\, \chE\, U = \int d^{12} z \, E\, \frac{Z}{\bar\nabla^4 Z} \, U.
\end{align}
However, in order for $\bar\nabla^4 Z$ to be both chiral and primary,
$Z$ must be of dimension \emph{zero}. This prevents us from adopting the
gauge $Z = 1$ and so no analogue of \eqref{eq_FtoDP} is possible\footnote{This
is usually understood by noting that the chiral projection operator in conventional
$\N=2$ superspace descriptions \eqref{eq_ChiralProjectionOperator} annihilates
any constant scalar field.},
at least within the superspace framework discussed here.\footnote{It \emph{is}
possible to construct something like \eqref{eq_FtoDP} if superspace is
augmented to include an internal $\mathbb CP^1$ manifold. The construction
is given in \cite{KT-M-normal}.}

\subsubsection{Component actions}
It remains to derive component actions from $\N=2$
actions in superspace. Since any $\N=2$ action over the full superspace
may be written as an integral over chiral superspace, the problem is,
as in $\N=1$, the derivation of the component form of a generic chiral action.
Within $\N=2$ conformal supergravity, this component action was constructed
originally via the superconformal tensor calculus \cite{sct_chiral}.
We give here a direct superspace construction of the same
result.\footnote{The construction for conventional $\N=2$ superspace
was given for the first time by M\"uller \cite{Muller2}.}

One begins with the action
\begin{align}
S = \int d^8\chz\, \chE\, U = \int d^{4}x\, e\, \Lag_c
\end{align}
where $U$ is a conformally primary chiral superfield of weight $(2,4)$;
our goal is to evaluate the component chiral Lagrangian $\Lag_c$. This can be done most
simply by first going to the specific $\theta$-dependent gauge, where
\begin{align}
E_{\bmu}{}^A = \delta_{\bmu}{}^A, \qquad
H_{\bmu}{}^{\ul a} = C_{\bmu \bnu} \theta^{\bnu} \,\bar \cW^{\ul a}.
\end{align}
Because of the constraints we have placed on $\bar \cW$, the spinor
covariant derivative in this gauge involves only three connection
terms,
\begin{align}\label{eq_spinorD}
\nabla_{\balpha} = \partial_{\balpha}
     - \frac{1}{2} \Omega_{\balpha \dbeta \dgamma} \bar M^{\dgamma \dbeta}
     - F_{\balpha}{}^b K_b
     - F_{\balpha \dbeta j} \bar S^{\dbeta j}.
\end{align}
In this gauge, $\chE = e \equiv \det\left(e_m{}^a\right)$ has nontrivial
$\theta$-dependence, so the component Lagrangian takes the form
\begin{align}
e \Lag_c &= \frac{1}{48} \partial^{ij} \partial_{ij} (e U) \eol
     &= \frac{1}{48} e \,\partial^{ij} \partial_{ij} U
     + \frac{1}{12} \partial^{\alpha i} e\, \partial_\alpha{}^j \partial_{ij} U
     + \frac{1}{16} \partial^{ij} e\, \partial_{ij} U
     - \frac{1}{16} \partial^{\alpha \beta} e\, \partial_{\alpha \beta} U
     \eol & \qquad
     + \frac{1}{12} \partial^{\alpha i} \partial_{ij} e\, \partial_\alpha{}^j U
     + \frac{1}{48} \partial^{ij} \partial_{ij} e \, U \label{eq_LagcWork}
\end{align}
where we have defined
\begin{gather}
\partial^{ij} \equiv \partial^{\beta i} \partial_\beta^j, \qquad
\partial_{\alpha \beta} \equiv \partial_{\alpha}{}^j \partial_{\beta j}.
\end{gather}
Because of the restricted form of \eqref{eq_spinorD}, one may replace
$\partial_\alpha{}^i \longrightarrow \nabla_\alpha{}^i$ in each of the above
terms. For the derivatives acting on $U$, this gives terms involving
covariant derivatives of $U$, which cannot be further simplified without
considering a specific form for $U$. However, the derivatives acting on
$e$ must be simplified further. One can proceed either by
explicitly constructing the $\theta$-expansion of $e$, or by making
use of the torsion relations recursively, which amounts to the same.
We take the second approach since it is easier in application.

There are two relations which we will need. The first is
\begin{align*}
\nabla_\alpha{}^i e = e \, \nabla_\alpha{}^i e_m{}^b e_b{}^m
     = e \, T_\alpha{}^i{}_m{}^b \,e_b{}^m
     = i e\, (\sigma^b \bpsi_b{}^{i})_\alpha.
\end{align*}
The second is
\begin{align*}
\nabla_\alpha{}^i \bpsi_{c \dbeta}{}^j &=
     (\nabla_\alpha{}^i e_c{}^m) \bpsi_{m \dbeta}{}^j
     + e_c{}^m \nabla_\alpha{}^i \bpsi_{m \dbeta}{}^j \eol
     &= -e_c{}^n T_{\alpha}{}^i{}_{n}{}^b \,\bpsi_{b \dbeta}{}^j 
     + 2 e_c{}^m T_\alpha{}^i{}_m{}_\dbeta{}^j \eol
     &= -i (\sigma^b \bar\psi_c{}^i)_\alpha \, \bar\psi_b{}_\dbeta{}^j
     + i \eps^{ij} \sigma_c{}_{\alpha \dgamma} \bar W^{\dgamma}{}_\dbeta.
\end{align*}
Applying these relations repeatedly allows us to evaluate all of the
required spinor derivatives of $e$:
\begin{align*}
\nabla^{ij} e &= 4 \,e \, (\bar\psi_m{}^i \bsigma^{mn} \bar\psi_n{}^j) \\
\nabla_{\alpha \beta} e &= -2 \,e\, (\bar\psi_m \bar\psi_{n}) (\sigma^{mn})_{\alpha \beta} \\
\nabla_\beta{}^i \nabla^{\beta \alpha} e &= -\nabla^{\alpha}{}_j \nabla^{ji} e =
     -3 \,e\, \eps^{abcd} (\bar\psi_a\bar\psi_b) (\sigma_c \bar\psi_d{}^i)^\alpha
     + 6 i \,e\, \bsigma_b^{\dgamma \alpha} \bar W_{\dgamma \dbeta} \,\bar\psi_b{}^{\dbeta i} \\
\nabla^{ij} \nabla_{ij} e &=
     -12i \,\eps^{abcd} \,(\bar\psi_a \bar\psi_b)(\bar\psi_c \bar\psi_d)
     + 48 \,(\bar\psi_a \bar\psi_b) (\bsigma^{ab})_{\dalpha \dbeta} \bar W^{\dbeta \dalpha}
     + 48 \,\bar W_{\dalpha \beta} \bar W^{\dbeta \alpha}
\end{align*}
where we have defined $(\bar\psi_a \bar\psi_b) \equiv \bar\psi_a{}^j \bar\psi_{b j}$.
Applying these results to \eqref{eq_LagcWork} gives
\begin{align}\label{eq_Lagc}
\Lag_c &= \nabla^4 U
     - \frac{i}{12} (\bar \psi_m{}^i \bsigma^m)^\alpha \nabla_\alpha{}^j \nabla_{ij} U
     \eol & \qquad
     + \frac{1}{4} (\bar\psi_m{}^i \bsigma^{mn} \bar\psi_n{}^j) \nabla_{ij} U
     + \frac{1}{8} (\bar\psi_m \bar\psi_{n})( \sigma^{mn})^{\alpha \beta} \nabla_{\alpha \beta} U
     \eol & \qquad
     + \frac{1}{4} \eps^{mnpq} (\bar\psi_{m} \bar\psi_{n})
          (\bar\psi_p{}^i \bsigma_q)^\alpha \nabla_{\alpha i} U
     + \frac{i}{2} \bar\psi_{b\dgamma}{}^j \,\bar W^\dgamma{}_\dbeta
          \,\bsigma_b^{\dbeta \alpha} \nabla_{\alpha j} \,U
     \eol & \qquad
     - \frac{i}{4} \eps^{mnpq} (\bar\psi_m \bar\psi_n) (\bar\psi_p \bar\psi_q) U
     + (\bar\psi_m \bar\psi_n) (\bsigma^{mn})^{\dalpha \dbeta} \bar W_{\dalpha \dbeta} U
     + \bar W^{\dalpha \dbeta} \bar W_{\dalpha \dbeta} U.
\end{align}
The action constructed from this Lagrangian is automatically
$\N=2$ superconformally invariant.

\section{Component analysis}\label{Section3}
In section \ref{Section2}, we have derived the superspace geometry corresponding to
conformal supergravity in superspace in a manner which \emph{manifestly}
respects the superconformal symmetry. In order for this to be the correct
superspace theory, it must reduce in components to the $\N=2$ superconformal
tensor calculus, which reproduces the structure of $\N=2$ conformal supergravity.
The field content of $\N=2$ conformal supergravity -- the so-called
``Weyl multiplet'' -- has been known for a long time \cite{sct_rules, sct_structure}.
It consists first of a set of one-forms: the vierbein $e_m{}^a$, the gravitino
$\psi_m{}^\alpha{}_i$, the $\rm U(1)_R$ and $\rm SU(2)_R$ gauge fields
$A_m$ and $\phi_m{}^i{}_j$, and the dilatation gauge field $b_m$. The last of these,
$b_m$, transforms algebraically under special conformal transformations
and may be gauged away. In addition there are gauge connections associated
with the rest of the superconformal group: the spin connection $\omega_m{}^{ab}$
and the special conformal and superconformal connections
$f_m{}^a$ and $\phi_m{}^{\alpha i}$, which are algebraically constrained
to be functions of other fields.

The story so far is similar to the $\N=1$ picture; however, counting degrees
of freedom, one finds that the number of fermionic degrees of freedom do not match the
number of bosonic ones. One is led to introduce additional ``matter'' fields:
a fermion $\chi_{\alpha i}$, an antisymmetric tensor $W_{ab}$,
and a real scalar $D$. Taking these into account, one finds 24 bosonic and 24
fermionic degrees of freedom. This set of fields was first identified
in \cite{sct_rules, sct_structure, Bergshoeff:1980is}.\footnotemark
\footnotetext{The linearized $\rm SO(2)$ supergravity was constructed earlier in
\cite{Fradkin:1979cw} and \cite{de Wit:1979pq}. This model 
was identitifed in \cite{sct_rules} as the Weyl multiplet
coupled to a vector and a nonlinear compensator. It is worth noting that
the Weyl multiplet is usually coupled
to a vector compensator to generate the so-called ``minimal field representation''
with 32 bosonic and 32 fermionic components \cite{BS:minimal_multiplet}.
A second compensator is needed to give
sensible equations of motion.}

Our goal is to show that our construction in superspace reduces in components
to that of \cite{sct_rules, sct_chiral, sct_imptensor, sct_structure, sct_gaugematter, sct_lagrangians}.
The superfields available are the one-forms
associated with each of the generators of the superconformal algebra as
well as the superfield $W_{\alpha \beta}$ and its various spinorial
derivatives. The component one-forms of the superconformal tensor calculus
obviously come from projections of the corresponding superspace one-forms:
\begin{gather}
e_m{}^a \equiv E_m{}^a\lc, \quad
\psi_m{}^\alpha{}_i \equiv 2 E_m{}^\alpha{}_i \lc, \quad
\bpsi_m{}_\dalpha{}^i \equiv 2 E_m{}_\dalpha{}^i \lc \eol
A_m \equiv A_m\lc, \quad
b_m \equiv B_m\lc, \quad
\omega_m{}^{ab} \equiv \Omega_m{}^{ab} \lc, \quad
\phi_m{}^i{}_j \equiv \Phi_m{}^i{}_j\lc \eol
f_m{}^a \equiv F_m{}^a\lc, \quad
\phi_m{}^{\alpha i} \equiv 2 F_m{}^{\alpha i} \lc, \quad
\bphi_m{}_{\dalpha i} \equiv 2 F_m{}_{\dalpha i} \lc.
\end{gather}
In contrast, the ``matter'' fields of the $\N=2$ Weyl multiplet must be identified
with certain components of the superfield $W_{\alpha \beta}$ and its
conjugate.\footnote{This identification seems to have first been made at
the linearized level by Bergshoeff, de Roo, and de Wit \cite{Bergshoeff:1980is}
who dealt with the general case of $\N \leq 4$, extending the earlier groundbreaking
work of Ferrara and Zumino \cite{Ferrara:1977mv} for $\N=1$.}
The lowest components of each have the correct
transformation properties to correspond to the self-dual and anti-self-dual parts
of $W_{ba}$:
\begin{gather}
W_{ba} = W_{ba}^+ + W_{ba}^-, \qquad
W_{ba}^+ \equiv (\sigma_{ba})_\beta{}^\alpha W_\alpha{}^\beta \lc, \qquad
W_{ba}^- \equiv (\bsigma_{ba})^\dbeta{}_\dalpha \bar W^\dalpha{}_\dbeta \lc \\
\frac{i}{2} \eps_{dc}{}^{ba} W_{ba}^{\pm} = \pm W_{dc}^\pm.
\end{gather}
The fermion $\chi_{\alpha i}$ and its conjugate correspond to
\begin{align}
\chi_{\alpha i} = -\frac{1}{3} \nabla^\beta{}_i W_{\beta \alpha} \lc, \qquad
\bar\chi^{\dalpha i} = \frac{1}{3} \bar\nabla_\dbeta{}^i \bar W^{\dbeta \dalpha} \lc~.
\end{align}
Finally, the scalar $D$ is given by
\begin{align}
D = \frac{1}{12} \nabla^{\alpha \beta} W_{\beta \alpha} \lc =
     \frac{1}{12} \bar\nabla_{\dalpha \dbeta} \bar W^{\dbeta \dalpha}\lc~,
\end{align}
which is real by virtue of the Bianchi identity \eqref{eq_WBianchi}
obeyed by $W_{\alpha \beta}$.

The other components of $W_{\alpha \beta}$ must then correspond to functions of
the gauge and ``matter'' fields. This is
certainly possible; recall that \emph{all} components of the $\N=1$ conformal superspace
curvature $W_{\alpha \beta \gamma}$ are functions of the $\N=1$ conformal supergravity
connections. Furthermore, the constraints on the curvatures in superspace must lead to
the same curvature constraints found in the STC formalism when reduced to components.

To understand how the reduction to components occurs, we will briefly review the
case of globally supersymmetric Yang-Mills. At the component level, this theory
consists of the gauge connection $\mathbf A_m$, the gaugino ${\boldsymbol\lambda}_{\alpha i}$,
a scalar field $\boldsymbol \phi$, and an auxiliary isotriplet $\mathbf D^{ij}$,
each of which is in the adjoint of the gauge group; for example, $\mathbf A_m = A_m{}^{r} \mathbf T_r$
where $\mathbf T_r$ is the generator in the adjoint. Within superspace,
the theory consists of a gauge connection $\mathbf A_M$ whose field strength
is constrained in terms of a reduced chiral superfield $\mathbf W$.  Only the vector part of
the superfield $\mathbf A_M$ contributes at the component level, whereas
the rest of the Yang-Mills multiplet, the ``matter'' fields, are found within the
chiral superfield $\mathbf W$ via
\begin{align}
\boldsymbol \phi \propto \mathbf W \lc, \quad
\boldsymbol \lambda_{\alpha i} \propto \CD_{\alpha i} \mathbf W \lc, \quad
\mathbf D^{ij} \propto \CD^{ij} \mathbf W \lc= \BCD^{ij} \bar {\mathbf W}\lc,
\end{align}
with the precise coefficients depending on one's conventions.
There remains the component $\CD_{\alpha \beta} \mathbf W$ and its conjugate,
which are not independent fields. Rather, they are fixed by the
vector field strength constraint
\begin{align}\label{eq_YMF}
\mathbf F_{ba}= - \frac{1}{8} (\sigma_{ba})^{\beta \alpha} \CD_{\alpha\beta} \mathbf W\lc
     - \frac{1}{8} (\bsigma_{ba})^{\dbeta \dalpha} \BCD_{\dalpha\dbeta} \bar {\mathbf W}\lc
\end{align}
where
\begin{align}\label{eq_YMFdef}
\mathbf F_{ba} \equiv e_b{}^n e_a{}^m \mathbf F_{nm}
     = e_b{}^n e_a{}^m \left(\partial_{[n} \mathbf A_{m]} - [\mathbf A_n, \mathbf A_m]\right)
\end{align}

For supergravity, the constraint analogous to \eqref{eq_YMF} for the
field strength is
\begin{align}\label{eq_Fdef}
\cF_{ba} = - (\sigma_{ba})^{\beta \alpha} \cF_{\sym{\alpha\beta}}
     - (\bsigma_{ba})^{\dbeta \dalpha} \bar \cF_{\sym{\dalpha\dbeta}}
\end{align}
where in conformal superspace
\begin{align}\label{eq_Fconstraint}
\cF_{\sym{\alpha \beta}} \equiv \frac{1}{8} \{\nabla_{(\alpha}{}^k ,[\nabla_{\beta) k},\cW]\}, \qquad
\cF_{\sym{\dalpha \dbeta}} \equiv \frac{1}{8} \{\bar\nabla_{(\dalpha}{}^k ,[\bar\nabla_{\dbeta) k},\bar\cW]\}~.
\end{align}
The analogue of the definition \eqref{eq_YMFdef} is more complicated in supergravity.
One begins with
\begin{align}
\cF_{nm} &=
     E_n{}^b E_m{}^a \cF_{ba}
     + E_{[n}{}^{\bbeta} E_{m]}{}^a \cF_{\bbeta a}
     + E_{[n \dbbeta} E_{m]}{}^a \cF^{\dbbeta}{}_{a}
     \eol & \quad
     - E_n{}^\bbeta E_m{}^\balpha \cF_{\bbeta \balpha}
     - E_{n \dbbeta} E_{m \dbalpha} \cF^{\dbbeta \dbalpha}
     - E_{[n}{}^{\bbeta} E_{m]}{}_{\dbalpha} \cF_{\bbeta}{}^\dbalpha
\end{align}
where as before we use a condensed summation convention for the bold indices $\balpha$
and $\dbalpha$:
\[
\chi^\balpha \lambda_\balpha \equiv \chi^\alpha{}_i \lambda_\alpha{}^i, \qquad
\bar\chi_\dbalpha \bar\lambda^\dbalpha \equiv \bar\chi_\dalpha{}^i \bar\lambda^\dalpha{}_i~.
\]
Projecting to lowest components and solving for $\cF_{ba} \lc$ yields
\begin{align}\label{eq_Rdef}
\cF_{ba}\lc &= e_b{}^n e_a{}^m \cF_{nm}
     - \frac{1}{2} (\psi_{[b} \sigma_{a]} \bar \lambda)
     + \frac{1}{2} (\bpsi_{[b} \bsigma_{a]}\lambda)
     \eol & \quad
     - \frac{1}{2} (\psi_b \psi_{a}) \bar \cW\lc
     - \frac{1}{2} (\bpsi_b \bpsi_{a}) \cW\lc
     - \frac{i}{2} (\psi_{[b} \sigma^c \bpsi_{a]}) \nabla_c
\end{align}
where we have applied the gauge constraints of conformal superspace \eqref{eq_gaugeconstraints}
and made use of the suppressed index conventions
\begin{gather}
(\psi_{n} \sigma_{m} \bar \lambda) = \psi_n{}^\alpha{}_j \, \sigma_{m \alpha \dalpha} \,\bar \lambda^{\dalpha j} ,\qquad
(\bpsi_{n} \bsigma_{m} \lambda) = \psi_{n \dalpha}{}^j \,\bsigma_m^{\dalpha \alpha} \,\lambda_{\alpha j} \\
(\psi_n \psi_{m}) = \psi_n{}^{\alpha j} \psi_{m \alpha j}, \qquad
(\bpsi_n \bpsi_{m}) =\bpsi_{n \dalpha}{}^j \bpsi_m{}^\dalpha{}_j
\end{gather}
along with the definitions
\begin{gather}
\lambda_{\alpha j} \equiv \cW_{\alpha j} \lc = -\frac{i}{2} [\nabla_{\alpha j}, \cW] \lc, \qquad
\bar \lambda^{\dalpha j} \equiv \bar\cW^{\dalpha j} \lc = -\frac{i}{2} [\bar\nabla^{\dalpha j}, \bar \cW] \lc~.
\end{gather}

We may take the relations \eqref{eq_Rdef} as the \emph{definitions} of
$\cF_{ba}\lc$ in analogy to \eqref{eq_YMFdef}. These
correspond to the so-called covariantized curvatures of the superconformal tensor
calculus. The constraints imposed on $\cF_{ba}\lc$ from \eqref{eq_Fconstraint}
will then correspond to the constraints found by de Wit et al.

\subsection{Comparison to superconformal tensor calculus}
In order to make comparisons between our results and those of the superconformal
tensor calculus, some conversions of notation are necessary. For definiteness, we use the
more recent conventions of \cite{LopesCardoso:2000qm,deWit:2006gn, deWit:2010za}.
The primary difference is that de Wit et al. make use of four component
spinors. Their Lorentz tensor conventions can be
exchanged for ours by the replacements
\begin{align}
\gamma^a \rightarrow
\left(\begin{array}{cc}
0 & i \sigma^a \\
i \bsigma^a & 0
\end{array}\right), \quad
\gamma^{ab} \rightarrow 
\left(\begin{array}{cc}
-2 \sigma^{ab} & 0 \\
0 & -2 \bsigma^{ab}
\end{array}\right),\quad
\eps_{abcd} \rightarrow -i \eps_{abcd}.
\end{align}
Second, their convention for $\rm SU(2)$ indices is opposite ours.
Lowered indices must be raised and vice-versa.
Third, there are some differing normalization
conventions for several of the gauge fields. Exchanging their
notation for ours necessitates making the replacements
\begin{subequations}
\begin{align}
A_m  &\rightarrow -2 A_m \\
\mathcal V_m{}^i{}_j &\rightarrow + 2 \phi_m{}_i{}^j \\
\phi_{m \alpha j} &\rightarrow -2 \phi_{m \alpha}{}^j \\
f_m{}^a &\rightarrow -2 f_m{}^a.
\end{align}
\end{subequations}
Additionally, the antisymmetric two-form of the Weyl multiplet
in STC carries antisymmetric $\rm SU(2)$ indices (which is natural for
$\N \geq 2$), whereas we use a notation specialized to $\N=2$. They
are related by
\begin{align}
T_{ab}{}^{ij} &\rightarrow -2 \eps_{ij} W_{ab}^+.
\end{align}
Finally, there are two major differences in the definition of curvatures.
The overall normalizations for $R(Q)_{cb}{}^{i}$ and $R(S)_{cb \,i}$ 
in STC differ from the corresponding curvatures in our approach
by a factor of two. That is,
\begin{align*}
 R(Q)_{cb}{}^i &= e_c{}^n e_b{}^m \CD_{[n} \psi_{m]}{}^i + \ldots \\
 R(S)_{cb \,i} &= e_c{}^n e_b{}^m \CD_{[n} \phi_{m] i} + \ldots
\end{align*}
whereas we have
\begin{align*}
T_{cb}{}^\alpha{}_i = \frac{1}{2} e_c{}^n e_b{}^m \CD_{[n} \psi_{m]}{}^\alpha{}_i + \ldots \\
R(S)_{cb}{}^{\alpha i} = \frac{1}{2} e_c{}^n e_b{}^m \CD_{[n} \phi_{m]}{}^{\alpha i} + \ldots.
\end{align*}
Taking these definitions as well as the gauge field normalizations into account,
curvatures in the STC approach may be related to our component curvatures via the
replacements
\begin{subequations}
\begin{align}
R(P)_{cb}{}^a &\rightarrow T_{cb}{}^a \\
R(Q)_{cb}{}^{\alpha i} &\rightarrow 2 T_{cb}{}^\alpha{}_i \\
R(M)_{dc}{}^{ba} &\rightarrow R(M)_{dc}{}^{ba} \\
R(A)_{ba} &\rightarrow -2 R(\A)_{ba} \\
R(D)_{ba} &\rightarrow R(\D)_{ba} \\
R(\mathcal V)_{ba}{}^i{}_j &\rightarrow 2 R(I)_{ba}{}_i{}^j \\
R(K)_{cb}{}^a &\rightarrow -2 R(K)_{cb}{}^a \\
R(S)_{cb}{}^\alpha{}_i &\rightarrow -4 R(S)_{cb}{}^{\alpha i}.
\end{align}
\end{subequations}

The constraint structure of the superconformal tensor calculus can be briefly
summarized in the following way. One introduces connections and
curvatures for the entire superconformal algebra. The curvatures are then
``covariantized'' using the matter fields of the $\N=2$ Weyl multiplet
and the covariantized curvatures are constrained by three relations,
which in our notation read
\begin{gather}
T_{cb}{}^a = 0 \\
T_{\gamma \dgamma\, \beta \dbeta\, }{}^\beta{}_i =
     -\frac{3}{2} \eps_{\dgamma \dbeta} \chi_{\gamma i} \\
R^c{}_{bca} = R(\D)_{ba} + \frac{3a}{2} \eta_{ba} D - \eta^{cd}\, W_{ac}^+ W_{bd}^-
\end{gather}
These constraints are \emph{algebraic} when describing the components; they
fix the spin connection $\omega_m{}^{ab}$, the special superconformal
connection $\phi_m{}^{\alpha i}$, and the special conformal connection $f_m{}^a$,
respectively. The parameter $a$ in the last expression is purely a matter of convention;
it has no effect on the physics, merely adding a term to the connection $f_m{}^a$
and simultaneously removing it elsewhere in any given action.\footnote{The choice $a=1$ is frequently
made, especially in recent work \cite{LopesCardoso:2000qm, deWit:2006gn, deWit:2010za}.}
It turns out that our approach will correspond to the choice $a=-2$.

It is easy to see why the choice $a=-2$ has been made. In the commutator
$[\delta_Q(\xi), \delta_{\bar Q}(\bar\xi)]$ constructed from the tensor calculus,
one finds \cite{sct_chiral}
\begin{align}
[\delta_Q(\xi), \delta_{\bar Q}(\bar\xi)] = -2i \,\delta_P(\xi \sigma^a\bar\xi)
     - \frac{i}{4} (a+2) \delta_K(\xi \sigma^a \bar\xi\, D).
\end{align}
This can be understood as arising from the superspace anticommutator
\begin{align}
\{\nabla_\alpha{}^i, \bar\nabla_{\dalpha j}\} = -2i \delta^i_j\, \nabla_{\alpha \dalpha}
     - \frac{i}{48} \delta^i_j \, (a+2) \nabla^{\gamma \beta} W_{\gamma \beta} \, K_{\alpha \dalpha}
\end{align}
Our choice $a=-2$ clearly simplifies the algebra, but any other value is equally
allowed; it simply imposes a conventional constraint on the curvature
\begin{align}
R(K)_\balpha{}^{\dbbeta\, c} = \frac{1}{96} (a+2)
     \,T_\balpha{}^{\dbbeta\, c}\, \nabla^{\gamma \beta} W_{\gamma \beta}.
\end{align}

The component curvatures have a quite intricate structure, and identically obey
a number of component Bianchi identities. We will evaluate each of their component
forms and analyze the corresponding constraints imposed.

\subsection{Torsion analysis}
We begin by analyzing the components of the torsion tensor.
Recall that $T_{nm}{}^a$ is given in lowest components by
\begin{align}
T_{nm}{}^a = \partial_{[n} e_{m]}{}^a + \omega_{[nm]}{}^a + b_{[n} e_{m]}{}^a.
\end{align}
Using the other components of the superspace torsion tensor, this can be
covariantized to
\begin{align}
T_{cb}{}^a = e_c{}^n e_b{}^m \partial_{[n} e_{m]}{}^a + \omega_{[cb]}{}^a + b_{[c} \delta_{b]}{}^a
     - \frac{i}{2} (\psi_{[c} \sigma^a \bpsi_{b]}).
\end{align}
This quantity is constrained to vanish,
\begin{align}
T_{cb}{}^a = 0,
\end{align}
which allows the spin connection $\omega_m{}^{ab}$ to be solved
in terms of the vierbein and the gravitino:
\begin{align}
\omega_{mnp} = e_n{}^a e_p{}^b \omega_{m a b} &=
     \frac{1}{2} ({e_m}^a \partial_n e_{pa} - {e_p}^a \partial_m e_{na}
     - {e_n}^a \partial_p e_{ma})
     \eol & \quad
     - \frac{1}{2} ({e_m}^a \partial_p e_{na} - {e_n}^a \partial_m e_{pa}
     - {e_p}^a \partial_n e_{ma})
     \eol & \quad
     + \frac{i}{4} (\psi_p \, \sigma_m \bpsi_n - \psi_m \sigma_n \bpsi_p
          - \psi_n\sigma_p \bpsi_m)
     \eol & \quad
     - \frac{i}{4} (\psi_n \, \sigma_m \bpsi_p - \psi_m \sigma_p \bpsi_n
          - \psi_p\sigma_n \bpsi_m)
     \eol & \quad
     + b_n g_{pm} - b_p g_{nm}.
\end{align}
This agrees with the conventional definition in supergravity, except for the
explicit appearance of the dilatation connection $b_m$.

\subsection{Gravitino torsion analysis}
The gravitino torsion tensor is given by
\begin{align}\label{eq_gravT1}
T_{nm}{}^\alpha{}_i =
     \frac{1}{2} \CD_{[n} \psi_{m]}{}^\alpha{}_i
     + \frac{i}{2} (\bphi_{[n i} \,\bsigma_{m]})^{\alpha},
\end{align}
where
\begin{align}
\CD_n \psi_m{}^\alpha{}_i &\equiv \partial_n \psi_m{}^\alpha{}_i
     + \psi_{m}{}^\beta{}_i \,\omega_{n \beta}{}^\alpha
     - i A_n \psi_m{}^\alpha{}_i
     + \phi_n{}_i{}^j \psi_m{}^\alpha{}_j
     + \frac{1}{2} b_n \psi_m{}^\alpha{}_i.
\end{align}
We have introduced the derivative $\CD$, which contains the Lorentz, $\rm U(2)_R$,
and dilatation connections. The gravitino torsion may be covariantized to
\begin{align}\label{eq_gravT2}
T_{cb}{}^\alpha{}_i
     &= \frac{1}{2} \psi_{cb}{}^\alpha{}_i
     + \frac{i}{2} (\bphi_{[c i} \,\bsigma_{b]})^{\alpha}
     + \frac{i}{8} \eps_{ik} (\bpsi_{[c}{}^k \bsigma_{b]} \sigma^{fd})^{\dgamma \alpha} W_{fd}^+ \\
\psi_{cb}{}^\alpha{}_i &= e_c{}^n e_c{}^m \psi_{nm}{}^\alpha{}_i \equiv
     e_c{}^n e_c{}^m \CD_{[n} \psi_{m]}{}^\alpha{}_i.
\end{align}
From superspace we have the relation
\begin{align}\label{eq_gravTC1}
T_{\gamma \dgamma\, \beta \dbeta\, }{}^\alpha{}_i =
     \frac{1}{2} \eps_{\dgamma \dbeta} \nabla^\alpha{}_i W_{\gamma \beta} \lc
\end{align}
which implies the constraint
\begin{align}\label{eq_gravTC2}
T_{\gamma \dgamma\, \beta \dbeta\, }{}^\beta{}_i  =
     -\frac{3}{2} \eps_{\dgamma \dbeta} \chi_{\gamma i}.
\end{align}
This allows the determination of $\bphi_{m\, \dalpha j}$:
\begin{align}
\bphi_{\beta \dbeta \, \dalpha j} &=
     -\frac{i}{6} \CD_\dbeta{}^\phi \psi_{\phi \dalpha \,\beta j}
     - \frac{i}{3} \CD_\dalpha{}^\phi \psi_{\phi \dbeta \,\beta j}
     - \frac{i}{6} \eps_{\dbeta \dalpha} \CD_\beta{}^\dphi \psi_{\phi \dphi\,}{}^\phi{}_j
     \eol & \quad
     - \frac{1}{6} W_\beta{}^\phi \bpsi_{\phi \dbeta \dalpha j}
     - \frac{1}{3} W_\beta{}^\phi \bpsi_{\phi \dalpha \dbeta j}
     + \frac{i}{6} \eps_{\dbeta \dalpha} \nabla^\phi{}_j W_{\phi \beta}\lc
\end{align}
which may equivalently be written
\begin{align}
\bphi_{m}{}^\dalpha{}_j &=
     \frac{i}{2} \left(\bsigma^{pn} \bsigma_m - \frac{1}{3} \bsigma_m \sigma^{pn} \right)^{\dalpha \beta}
          \left(\CD_p \psi_{n\, \beta j} -
               \frac{i}{4} W_{ab}^+ (\sigma^{ab} \sigma_p \bar\psi_{n j})_\beta \right)
     - \frac{i}{4} (\bsigma_{m} \chi_{j})^\dalpha \eol
     &= \frac{i}{2} \left(\bsigma^{pn} \bsigma_m - \frac{1}{3} \bsigma_m \sigma^{pn} \right)^{\dalpha \beta}
          \CD_p \psi_{n\, \beta j}
     - \frac{1}{3} W_{mn}^+ \bar\psi^n{}_{\dalpha j}
     - \frac{1}{3} W_{mb}^+ (\bsigma^{bn} \bar\psi_{n\, j})_{\dalpha}
     - \frac{i}{4} (\bsigma_{m} \chi_{j})^\dalpha .
\end{align}
Applying this result to \eqref{eq_gravTC1} determines the spin-3/2 part of the
spinor derivative of $W_{\alpha \beta}$
\begin{align}
\sum_{(\alpha \beta \gamma)} \nabla_\gamma{}^j W_{\beta \alpha}\lc
     = \sum_{(\alpha \beta \gamma)} \left(\CD_\beta{}^\dphi \psi_{\dphi \alpha \,\gamma}{}^j
     - i \bpsi_{\beta \dphi}{}^{\dphi j} W_{\alpha \gamma}\right).
\end{align}
The sum is over all permutations of the indices.
This relation is analogous to the definition of $W_{\alpha \beta \gamma}$ in
$\N=1$ supergravity.

For the conjugate formulae, one finds
\begin{align}
\phi_{\beta \dbeta\,}{}_\alpha{}^j &=
     + \frac{i}{6} \CD_\beta{}^\dphi \bpsi_{\alpha \dphi\,}{}_\dbeta{}^j
     + \frac{i}{3} \CD_\alpha{}^\dphi \bpsi_{\beta \dphi\,}{}_\dbeta{}^j
     + \frac{i}{6} \eps_{\beta \alpha} \CD_\dbeta{}^\phi \psi_{\phi \dphi}{}^{\dphi j}
     \eol & \quad
     - \frac{1}{6} \bar W_\dbeta{}^\dphi \psi_{\beta \dphi\,}{}_\alpha{}^j
     - \frac{1}{3} \bar W_\dbeta{}^\dphi \psi_{\alpha \dphi\, \beta}{}^j
     - \frac{i}{6} \eps_{\beta \alpha} \bar\nabla^{\dphi j} \bar W_{\dphi \dbeta}\lc
\end{align}
equivalently written as
\begin{align}
\phi_{m}{}_\alpha{}^j &=
     \frac{i}{2} \left(\sigma^{pn} \sigma_m - \frac{1}{3} \sigma_m \bsigma^{pn} \right)_{\alpha \dbeta}
          \left(\CD_p \bpsi_{n}{}^{\dbeta j} +
               \frac{i}{4} W_{ab}^- (\bsigma^{ab} \bsigma_p \psi_{n}{}^j)^\dbeta \right)
     - \frac{i}{4} (\sigma_{m} \bar\chi^{j})_\alpha \eol
     &= \frac{i}{2} \left(\sigma^{pn} \sigma_m - \frac{1}{3} \sigma_m \bsigma^{pn} \right)_{\alpha \dbeta}
          \CD_p \bpsi_{n}{}^{\dbeta j}
     + \frac{1}{3} W_{mn}^- \psi^n{}_{\alpha}{}^j
     + \frac{1}{3} W_{mb}^- (\sigma^{bn} \psi_{n}{}^j)_{\alpha}               
     - \frac{i}{4} (\sigma_{m} \bar\chi^{j})_\alpha,
\end{align}
along with the relation
\begin{align}
\sum_{(\dalpha \dbeta \dgamma)} \bar\nabla_\dgamma{}^j \bar W_{\dbeta \dalpha}\lc
     &= \sum_{(\dalpha \dbeta \dgamma)} \left(\CD_\dbeta{}^\phi \bpsi_{\phi \dalpha\, \dgamma}{}^j
     + i \psi_{\phi \dbeta}{}^{\phi j} \bar W_{\dalpha \dgamma}\right).
\end{align}

\subsection{Dilatation and $\rm U(1)_R$ curvatures}
The dilatation and axial curvatures are given by
\begin{align}
R(\D)_{nm} &= \partial_{[n} b_{m]} + 2 f_{[nm]}
     + \frac{1}{2} \psi_{[n} \phi_{m]}
     + \frac{1}{2} \bpsi_{[n} \bphi_{m]} \\
R(\A)_{nm} &= \partial_{[n} A_{m]}
     - \frac{i}{4} \psi_{[n}\phi_{m]}
     + \frac{i}{4} \bpsi_{[n}\bphi_{m]}.
\end{align}
Their supercovariant forms are 
\begin{align}
R(\D)_{ba} &=
     e_b{}^n e_a{}^m \partial_{[n} b_{m]}
     + 2 f_{[ba]}
     + \frac{1}{2} \psi_{[b}\phi_{a]}
     + \frac{1}{2} \bpsi_{[b}\bphi_{a]}
     + \frac{3i}{8} (\psi_{[b} \sigma_{a]} \bar\chi)
     + \frac{3i}{8} (\bpsi_{[b} \bsigma_{a]} \chi) \\
R(\A)_{ba} &= e_b{}^n e_a{}^m \partial_{[n} A_{m]}
     - \frac{i}{4} \psi_{[b}\phi_{a]}
     + \frac{i}{4} \bpsi_{[b}\bphi_{a]}
     - \frac{3}{16} (\psi_{[b} \sigma_{a]} \bar\chi)
     + \frac{3}{16} (\bpsi_{[b} \bsigma_{a]} \chi)
\end{align}
where we have used the suppressed index convention
\begin{align}
(\psi_{b} \sigma_{a} \bar\chi) = (\psi_{b\, j} \sigma_{a} \bar\chi^j), \qquad
(\bpsi_{b} \bsigma_{a} \chi) = (\bpsi_{b}{}^j \sigma_{a} \chi_j)
\end{align}
with both sets of isospin indices in their natural positions.

However, these are constrained by
\begin{gather}
R(\D)_{\sym{\beta\alpha}} = 2i R(\A)_{\sym{\beta\alpha}} =
     \frac{1}{16} \nabla_{\beta}{}^\gamma W_{\gamma \alpha}\lc
     + \frac{1}{16} \nabla_{\alpha}{}^\gamma W_{\gamma \beta}\lc \\
R(\D)_{\sym{\dbeta\dalpha}} = -2i R(\A)_{\sym{\dbeta\dalpha}} =
     \frac{1}{16} \bar\nabla_{\dbeta \dgamma} \bar W^{\dgamma}{}_\dalpha \lc
     + \frac{1}{16} \bar\nabla_{\dalpha \dgamma} \bar W^{\dgamma}{}_\dbeta \lc.
\end{gather}
These relations accomplish two things: they fix
$\nabla_{(\beta}{}^\gamma W_{\alpha) \gamma}\lc$ and also require
\begin{align}
R(\D)_{dc} = -\eps_{dc}{}^{ba} R(\A)_{ba}.
\end{align}
This latter condition in principle imposes a relation on $f_{[dc]}$; but,
as we will shortly see, $f_m{}^a$ is \emph{entirely} determined by a constraint
on the Lorentz curvature just as in the superconformal tensor calculus.
The above result is really a consistency condition and is identically
satisfied as in the superconformal tensor calculus.

\subsection{Lorentz curvature}
The Lorentz curvature is given by
\begin{align}
R_{nm}{}^{ba} = 
     \partial_{[n} \omega_{m]}{}^{ba}
     - \omega_{[n}{}^{bc} \omega_{m]c}{}^a
     + 2 e_{[n}{}^{[b} f_{m]}{}^{a]}
     + (\psi_{[n} \sigma^{ba} \phi_{m]})
     + (\bpsi_{[n} \bsigma^{ba} \bphi_{m]})
\end{align}
where we have used the shorthand
\begin{align}
(\psi_{n} \sigma^{ba} \phi_{m}) = (\psi_{n\, j} \sigma^{ba} \phi_{m}{}^j), \qquad
(\bpsi_{n} \bsigma^{ba} \bphi_{m}) = (\bpsi_{n}{}^j \bsigma^{ba} \bphi_{m\, j})
\end{align}
with the isospin indices in their natural positions.

It will be useful to introduce the symbol $\poin R_{nm}{}^{ba}$ to
correspond to the Poincar\'e version of the Lorentz curvature; that is,
it equals the above expression without $f_m{}^a$ and $\phi_m{}^{\alpha i}$.
The covariantized Lorentz curvature is
\begin{align}
R_{dc}{}^{ba} &= 
     \poin R_{dc}{}^{ba}
     + 2 \delta_{[d}{}^{[b} f_{c]}{}^{a]}
     + (\psi_{[d} \sigma^{ba} \phi_{c]})
     + (\bpsi_{[d} \bsigma^{ba} \bphi_{c]})
     \eol & \quad
     - \frac{1}{2} (\psi_{[d} \sigma_{c]} \bar\lambda^{ba})
     + \frac{1}{2} (\bpsi_{[d} \bsigma_{c]} \lambda^{ba})
     - \frac{1}{2} \psi_d \psi_c W^{ba -}
     - \frac{1}{2} \bpsi_d \bpsi_c W^{ba +}
\end{align}
where we have used
\begin{align}
(\psi_d \psi_c) = (\psi_d{}^j \psi_{c j}), \qquad
(\bpsi_d \bpsi_c) = (\bpsi_d{}^j \bpsi_{c j}).
\end{align}
The fermionic combination $\lambda_{\alpha j}{}^{ba}$ is equal to a certain
combination of spinor derivatives of $W_{\alpha \beta}$; it may equivalently
be written
\begin{align}
\lambda_{\alpha j}{}^{ba} = i (\sigma^{dc} \sigma^{ba} T_{dc\, j}).
\end{align}
Making use of the constraint \eqref{eq_gravTC2}, this can be rewritten as
\begin{align}
\lambda_{\alpha j}{}^{ba} = -2i T^{ba}{}_{\alpha j} + \frac{3i}{2} (\sigma^{ba} \chi_j)_\alpha,
\end{align}
which allows the final form of the covariantized Lorentz curvature to be written
\begin{align}
R_{dc}{}^{ba} &= 
     \poin R_{dc}{}^{ba}
     + 2 \delta_{[d}{}^{[b} f_{c]}{}^{a]}
     + (\psi_{[d} \sigma^{ba} \phi_{c]})
     + (\bpsi_{[d} \bsigma^{ba} \bphi_{c]})
     - i (\psi_{[d} \sigma_{c]} \bar T^{ba})
     - i (\bpsi_{[d} \bsigma_{c]} T^{ba})
     \eol & \quad
     + \frac{3i}{4} (\psi_{[d} \sigma_{c]} \bsigma^{ba} \bar\chi)
     + \frac{3i}{4} (\bpsi_{[d} \bsigma_{c]} \sigma^{ba} \chi)
     - \frac{1}{2} \psi_d \psi_c W^{ba -}
     - \frac{1}{2} \bpsi_d \bpsi_c W^{ba +}. \label{eq_LCurv}
\end{align}
Because $R_{dcba}$, when written in spinor form, involves the totally symmetric
combination $\nabla_{(\delta \gamma} W_{\beta \alpha)}$, the above 
relation determines the lowest component of this superfield.

However, the lower spin parts of $R_{dcba}$ are constrained. Contracting
indices to construct the Ricci tensor, we find
\begin{align}
\cR_{(\gamma \dgamma) (\alpha \dalpha)} =
     \frac{1}{4} \eps_{\dgamma \dalpha} \nabla^{\phi}{}_{(\gamma} W_{\alpha) \phi} \lc
     - \frac{1}{4} \eps_{\gamma \alpha} \bar\nabla_{\dphi (\dgamma} \bar W_{\dalpha)}{}^\dphi \lc
     + \frac{1}{2} \eps_{\gamma \alpha} \eps_{\dgamma \dalpha} (\nabla^{\delta \beta} W_{\beta \delta}) \lc
     + 2 W_{\gamma \alpha} W_{\dgamma \dalpha}.
\end{align}
All of the terms on the right hand side have already been determined.
In vector form, this relation can be written
\begin{align}
\cR_{ba} = R(\D)_{ba} - 3\eta_{ba} D - \eta^{cd}\, W_{ac}^+ W_{bd}^-
\end{align}
which is precisely the constraint we wanted to show for the choice $a=-2$.
This allows the determination of $f_m{}^a$ as in the superconformal tensor
calculus. Usually one needs only $f_a{}^a$, which is given by
\begin{align}
f_a{}^a &= -D -\frac{1}{12} \poin \cR
     - \frac{1}{24} \eps^{abcd} (\bpsi_a \bsigma_b \CD_c \psi_d)
     + \frac{1}{24} \eps^{abcd} (\psi_a \sigma_b \CD_c \bpsi_d)
     \eol & \quad
     - \frac{i}{8} (\psi_b \sigma^b \bar \chi)
     - \frac{i}{8} (\bpsi_b \bsigma^b \chi)
     + \frac{1}{12} W^{ab +} (\bpsi_a \bpsi_b)
     + \frac{1}{12} W^{ab-} (\psi_a \psi_b)
\end{align}
where we have defined
\begin{align}
(\psi_a \sigma_b \CD_c \bpsi_d) = (\psi_{a j} \sigma_b \CD_c \bpsi_d{}^j)\,, \qquad
(\bpsi_a \bsigma_b \CD_c \psi_d) = (\bpsi_{a}{}^j \bsigma_b \CD_c \psi_{d j})
\end{align}
with the isospin indices in their natural positions.

\subsection{Isospin curvature}
The isospin curvature is given by
\begin{align}
R(I)_{nm \, ij} = \partial_{[n} \phi_{m]\, ij}
     + \phi_{[n (i}{}^k \phi_{m] j) k}
     - \psi_{[n (i} \phi_{m] j)}
     + \bpsi_{[n (i} \bphi_{m] j)}.
\end{align}
Its covariantized form is
\begin{align}
R(I)_{ba \, ij} &= e_b{}^n e_a{}^m \partial_{[n} \phi_{m]\, ij}
     + \phi_{[b (i}{}^k \phi_{a] j) k}
     - \psi_{[b (i} \phi_{a] j)}
     + \bpsi_{[b (i} \bphi_{a] j)}
     \eol & \quad
     - \frac{3i}{4} \psi_{[b (i} \sigma_{a]} \bar\chi_{j)}
     + \frac{3i}{4} \bpsi_{[b (i} \bsigma_{a]} \chi_{j)}.
\end{align}
This is constrained by the superspace structure to obey
\begin{align}
R(I)_{\beta \dbeta\, \alpha \dalpha\, ij}
     = -\frac{1}{4} \eps_{\dbeta \dalpha} \nabla_{ij} W_{\beta \alpha}\lc
     +\frac{1}{4} \eps_{\beta \alpha} \bar\nabla_{ij} \bar W_{\dbeta \dalpha}\lc.
\end{align}
This superspace constraint serves only to define $\nabla_{ij} W_{\beta \alpha}\lc$
and its conjugate; there is no constraint on the component curvature.

\subsection{Special superconformal curvature}
The special superconformal curvature is given by
\begin{align}
R(S)_{nm}{}^{\alpha j} &=
     \frac{1}{2} \CD_{[n} \phi_{m]}{}^{\alpha j}
     - \frac{i}{2} (\bpsi_{[n}{}^j \bsigma_c)^\alpha\, f_{m]}{}^c.
\end{align}
Its covariantized form is rather complicated. We begin with
the defining relation,
\begin{align}
R(S)_{ba}{}^{\alpha j} &=
     e_b{}^n e_a{}^m R(S)_{nm}{}^{\alpha j}
     - \frac{1}{2} (\psi_{[b} \sigma_{a]} \bar\lambda^{\alpha j})
     + \frac{1}{2} (\bpsi_{[b} \bsigma_{a]} \lambda^{\alpha j})
     + \frac{3}{8} (\bpsi_{[b} \bpsi_{a]}) \chi^{\alpha j}
\end{align}
where we have made use of the relations
\begin{align}
\cW(S)^{\alpha j} \lc= \frac{1}{4} \nabla^{\beta j} W_{\beta}{}^\alpha \lc = -\frac{3}{4} \chi^{\alpha j}, \quad
\bar\cW(S)^{\alpha j} = 0.
\end{align}
The expressions for $\lambda$ and $\bar\lambda$ are rather complicated.
The simpler is $\bar\lambda$, which can be written
\begin{align}
\bar\lambda^{\dbeta k\, \alpha j} = \frac{1}{4} \nabla_{\dphi}{}^\alpha \bar W^{\dphi \dbeta}\lc \, \eps^{kj} 
     = \frac{1}{8} \nabla_c \bar W_{ba}{}^- (\bsigma_{ba} \bsigma_c)^{\dbeta \alpha} \eps^{kj}.
\end{align}
The expression for $\lambda$, on the other hand, is more intricate:
\begin{align}
\lambda_{\beta k}{}^{\alpha j} &=
     - \frac{i}{8} \nabla_{\beta k} \nabla^{\phi j} W_\phi{}^\alpha\lc.
\end{align}
Decomposing its Lorentz and isospin structures into irreducible representations,
we find
\begin{align}
\lambda_{\beta k}{}^{\alpha j} &=
     \frac{i}{16} \nabla_j{}^k W_\beta{}^\alpha\lc
     + \frac{i}{16} \eps^{\alpha \gamma} \nabla^\phi{}_{(\beta} W_{\gamma) \phi}\lc
     - \frac{i}{32} \nabla^{\delta \phi} W_{\delta \phi} \,\delta_k{}^j \delta_\beta{}^\alpha\lc.
\end{align}
Each of these structures has already been specified, leading to
\begin{align}
\lambda_{\beta k}{}^{\alpha j} &=
     \frac{i}{4} R(I)_{ba}{}_j{}^k (\sigma^{ba})_\beta{}^\alpha
     + \frac{1}{2} \delta_k^j\, R(\A)_{ba} (\sigma^{ba})_\beta{}^\alpha
     - \frac{3i}{8} D \,\delta_k{}^j \delta_\beta{}^\alpha.
\end{align}
This yields the covariantized form
\begin{align}
R(S)_{ba}{}^{\alpha j} &=
     \frac{1}{2} e_b{}^n e_a{}^m \CD_{[n} \phi_{m]}{}^{\alpha j}
     + \frac{i}{2} f_{[b}{}^c (\bpsi_{a]}{}^j \bsigma_c)^\alpha
     + \frac{1}{16} \nabla_f W_{dc}^- (\psi_{[b}{}^j \bsigma_{a]} \bsigma_{dc} \bsigma_f)^\alpha
     \eol & \quad
     + \frac{i}{8} R(I)_{dc\,}{}_k{}^j (\bpsi_{[b}{}^k \bsigma_{a]})^\alpha
     + \frac{1}{4} R(\A)_{dc}\, (\bpsi_{[b}{}^j \bsigma_{a]} \sigma^{dc})^\alpha 
     - \frac{3i}{16} D \, (\bpsi_{[b}{}^j \bsigma_{a]})^\alpha
     \eol & \quad
     + \frac{3}{8} (\bpsi_{[b} \bpsi_{a]}) \chi^{\alpha j}.
\end{align}
It should be emphasized that the above expression results from the
choice $a=-2$.

The constraints imposed by superspace are naturally written in spinor form.
We find first
\begin{align}
R(K)_{\sym{\dgamma \dbeta}}{}_{\,\alpha}{}^i &=
     - \frac{i}{16} \nabla_\alpha{}^\dphi \bar\nabla_\dgamma{}^j \bar W_{\dbeta \dphi}\lc
     - \frac{i}{16} \nabla_\alpha{}^\dphi \bar\nabla_\dbeta{}^j \bar W_{\dgamma \dphi}\lc
     - \frac{1}{8} \bar W_{\dgamma \dbeta}  \nabla^{\phi j} W_{\phi \alpha} \lc \eol
     &= \frac{i}{4} \nabla_\alpha{}^\dphi T_{\sym{\dphi \dgamma}}{}_\dbeta{}^j
     + \frac{i}{4} \nabla_\alpha{}^\dphi T_{\sym{\dphi \dbeta}}{}_\dgamma{}^j
     + \frac{3}{8} \bar W_{\dgamma \dbeta} \chi_\alpha{}^j.
\end{align}
This constraint is actually obeyed identically. It corresponds
to part of the spinor decomposition of eq. (A.13) of \cite{deWit:2010za}.
The other part of the curvature tensor is
\begin{align}\label{eq_temp3.73}
R(K)_{\sym{\gamma \beta}}{}_{ \,\alpha}{}^i &=
     \frac{1}{96}
          \left(\nabla^{ij} \nabla_{\gamma j} W_{\beta \alpha}\lc
               + \nabla^{ij} \nabla_{\beta j} W_{\gamma \alpha}\lc
          \right).
\end{align}
Here one must recall that $W_{\alpha \beta}$ is constrained by its Bianchi
identity \eqref{eq_WBianchi}. In particular, the Bianchi identity implies
\begin{align}
\nabla^{ij} \nabla^\gamma{}_j W_{\gamma \alpha} =
     -6i \nabla_{\alpha \dbeta} \bar\nabla_\dgamma{}^i \bar W^{\dgamma \dbeta}
\end{align}
so the spin-1/2 part of \eqref{eq_temp3.73} is constrained; the spin-3/2 part
serves to define the totally symmetric
$\nabla^{ij} \nabla_{(\gamma j} W_{\beta \alpha)}\lc$. Note that this
combination is the last remaining component of the $W_{\alpha \beta}$ which
requires definition; all other components have been fixed or are
related to already defined objects by the Bianchi identity.

Taking just the spin-1/2 part of the curvature, we find
\begin{align}
R(S)_{\alpha \beta}{}^{\beta j} = \frac{3i}{16} \nabla_{\beta \dphi}{}\bar\chi^{\dphi j}.
\end{align}
This is a constraint, but it corresponds to the rest of the spinor decomposition
of eq. (A.13) of \cite{deWit:2010za}. Thus the constraints on the special
conformal curvature are \emph{exactly} as in the superconformal tensor
calculus, and so they are satisfied identically.

\subsection{Special conformal curvature}
The special conformal curvature is given by
\begin{align}
R(K)_{nm}{}^a = \CD_{[n} f_{m]}{}^a + \frac{i}{2} (\phi_{[n} \sigma^a \bphi_{m]}).
\end{align}
Again we have suppressed isospin indices, taking
\begin{align}
 (\phi_{n} \sigma^a \bphi_{m}) = (\phi_{n}{}^j \sigma^a \bphi_{m\, j}) 
\end{align}
with the indices in their natural positions.

Its supercovariant form is a bit of an exercise in restoring $\sigma$ matrices
everywhere but otherwise is completely straightforward. The result may be written
\begin{align}
R(K)_{cb}{}^a &= e_c{}^n e_b{}^m \CD_{[n} f_{m]}{}^a + \frac{i}{2} (\phi_{[c} \sigma^a \bphi_{b]})
     + \frac{i}{2} (\psi_{[c j} \sigma_{b]} \nabla_d T^{da \,j})
     + \frac{i}{2} (\bpsi_{[c}{}^j \bsigma_{b]} \nabla_d T^{da}{}_j)
     \eol & \quad
     + \frac{1}{4} (\psi_c \psi_b) \nabla_d W^{-\, da}
     + \frac{1}{4} (\bpsi_c \bpsi_b) \nabla_d W^{+\, da}.
\end{align}

It turns out that the special conformal curvature is completely constrained
by the superspace structure to equal other objects; however, this is exactly
the same situation as in STC. The precise form of the curvature in terms of
superfields is given in \eqref{eq_KcurvApp}; we focus merely on highlighting
the self-dual part in spinor notation:
\begin{align}
R(K)_{\sym{\beta \alpha}}{}_{\,\gamma \dgamma}
     = -\frac{1}{2} \nabla_\dgamma{}^\phi R_{\sym{\phi \gamma} \sym{\beta \alpha}}
     - \sum_{\beta \alpha} \Big(
          \frac{3i}{4} \bar\chi_{\dgamma j} T_{\sym{\gamma\beta }}{}_\alpha{}^j
          + \frac{1}{8} \nabla_{\alpha \dphi} (\bar W^\dphi{}_{\dgamma} W_{\beta\gamma})
          + \frac{3}{8} \eps_{\alpha \gamma} \nabla_{\beta \dgamma} D \Big).
\end{align}
It is a straightforward (and tedious) exercise to show that this form is
implied by the relation (A.8) of \cite{deWit:2010za}, which
is an identity on the component fields.

\section{Reduction to $\rm SL(2,\mathbb C) \times \rm U(2)_R$}\label{Section4}
Howe's formulation of $\N=2$ conformal supergravity \cite{Howe} involved
the choice of structure group $\rm SL(2,\mathbb C) \times \rm U(2)_R$, with the
corresponding covariant derivative
\begin{align}\label{eq_PoinD}
\CD_A = E_A{}^M \left(\partial_M - \frac{1}{2} \Omega_M{}^{ba} M_{ab}
     - A_M \A - \Phi_M{}^i{}_j I^j{}_i\right)
\end{align}
with the algebra
\begin{align}
[\CD_A, \CD_B] = -\poin T_{AB}{}^C - \frac{1}{2} \poin R_{AB}{}^{cd} M_{dc}
     - \poin R_{AB}{}^i{}_j I^j{}_i - \poin R_{AB} \A.
\end{align}
In order for the component structure of the theory to correspond to
conformal supergravity, the torsion tensor must be constrained by
\begin{gather}
\poin T_{\balpha \bbeta}{}^C = \poin T^{\dbalpha \dbbeta C} = \poin T_{\balpha}{}^{\dbbeta \bgamma} = 
      \poin T_{\balpha}{}^{\dbbeta}{}_{\dbgamma} = 0 \eol
\poin T_\alpha{}^i{}^\dbeta{}_j{}^c = 2i \,\delta^i_j \,(\sigma^c)_\alpha{}^\dbeta  \eol
\poin T_{A b}{}^c = 0, \quad
\poin T_{\alpha \dalpha\, \beta}{}^j{}^\gamma{}_k = \frac{1}{2} \delta_\alpha{}^\gamma \,\poin T_{\phi \dalpha\, \beta}{}^j{}^\phi{}_k \label{eq_Tconstraints}
\end{gather}
The solution to these constraints is then uniquely given by Howe's
superspace formulation.\footnote{See \cite{KLRT_M_Howe} for a recent detailed review of
Howe's formulation of $\N=2$ conformal supergravity.} The superfield
content of this theory involves three complex
superfields\footnote{It should be noted that when comparing
our notation to \cite{KLRT_M_Grimm, KLRT_M_Howe}, we have an extra
sign appearing in the Lorentz decomposition of an antisymmetric
two-form \eqref{eq_LorentzDecomposition}. Thus while our
$\bar W_{\dalpha \dbeta}$ and $Y_{\alpha \beta}$ match
those of \cite{KLRT_M_Grimm, KLRT_M_Howe}, our $W_{\alpha \beta}$
and $\bar Y_{\dalpha \dbeta}$ differ by a sign.}
\begin{gather}
W_{\alpha \beta} = W_{\beta \alpha}, \quad
Y_{\alpha \beta} = Y_{\beta \alpha}, \quad
S_{ij} = S_{ji},
\end{gather}
and two real superfields
\begin{gather}
G_a, \quad
G_a{}^{ij} = G_a{}^{ji}
\end{gather}
along with a set of constraints,
\begin{subequations}
\begin{align}
\CD_\alpha{}^{(i} S^{jk)} &= 0, \quad
\BCD_\dalpha{}^{(i} S^{jk)} = i \CD^{\beta (i} G_{\beta \dalpha}{}^{jk)} \\
\CD_{(\alpha}^i Y_{\beta \gamma)} &= 0, \quad
\CD_\alpha{}^i S_{ij} + \CD^\beta{}_j Y_{\beta \alpha} = 0 \\
\CD_{(\alpha}^{(i} G_{\beta) \dbeta}{}^{jk)} &= 0 \\
\CD_\alpha{}^i G_{\beta \dbeta} &= -i \BCD_\dbeta{}^i Y_{\beta \alpha}
     + \frac{1}{12} \eps_{\alpha \beta} \BCD_{\dbeta j} S^{ji}
     - \frac{1}{4} \eps_{\alpha \beta} \BCD^{\dgamma i} \bar W_{\dgamma \dbeta}
     - \frac{i}{3} \eps_{\alpha \beta} \CD^\gamma{}_j G_{\gamma \dbeta}{}^{ji}.
\end{align}
\end{subequations}

What makes this formulation conformal is that the torsion constraints admit
a super-Weyl transformation involving a real unconstrained superfield parameter
$\Lambda$. In terms of this super-Weyl transformation, $W_{\alpha \beta}$ transforms 
homogeneously, but all the other fields above transform inhomogeneously:
\begin{subequations}\label{eq_HoweSW}
\begin{align}
\delta W_{\alpha \beta} &= \Lambda W_{\alpha \beta} \\
\delta Y_{\alpha \beta} &= \Lambda Y_{\alpha \beta} - \frac{1}{2} \CD_{\alpha \beta} \Lambda \\
\delta S_{ij} &= \Lambda S_{ij} - \frac{1}{2} \CD_{ij} \Lambda \\
\delta G_{\alpha \dalpha} &= \Lambda G_{\alpha \dalpha} - \frac{1}{8} [\CD_\alpha{}^k, \BCD_{\dalpha k}] \Lambda \\
\delta G_{\alpha \dalpha}{}^{ij} &= \Lambda G_{\alpha \dalpha}{}^{ij}
     + \frac{i}{4} [\CD_\alpha{}^{(i}, \BCD_{\dalpha}{}^{j)}] \Lambda.
\end{align}
\end{subequations}
It is clear that $W_{\alpha \beta}$ should be the field of the same name from conformal
superspace; the other fields should have an origin which sheds light on their
transformation properties and their occurence in the curvature and torsion
tensors.

\subsection{Conventional degauging}
Our structure group differs from the group $\rm SL(2,\mathbb C) \times \rm U(2)_R$
by the addition of the dilatation symmetry and the special conformal
generators $K_A$. As Howe's formulation has as an extra symmetry
of the constraints only the super-Weyl transformation, we must fix the
special conformal symmetry in the degauging procedure.

It turns out there is an obvious way of doing this, and it works the same way
for $\N=2$ as for $\N\leq 1$. Under a special conformal transmation with parameter
$\eps^A$, the dilatation gauge field $B_M$ transforms as
\begin{align}
\delta_K(\eps) B_M = -2 \eps^a E_{M a}
     + 2 \eps^{\alpha j} E_{M \alpha j}
     + 2 \eps_{\dalpha j} E_M{}^{\dalpha j}.
\end{align}
It is evidently possible that we may exhaust the $K$-gauge freedom via
the gauge choice $B_M=0$. This not only fixes the special conformal symmetry
but also removes the dilatation connection from all covariant derivatives.

However, the covariant derivatives still carry the special conformal
connection $F_M{}^A$. As the symmetry associated with this connection
has been fixed, it should no longer be considered as part of the covariant
derivative. We therefore rewrite
\begin{align}
\nabla_A = \CD_A - F_A{}^B K_B, \quad
F_A{}^B \equiv E_A{}^M F_M{}^B
\end{align}
where $\CD_A$ is the $\rm SL(2,\mathbb C) \times \rm U(2)_R$
covariant derivative \eqref{eq_PoinD}.

It is quite straightforward to work out how the Poincar\'e curvatures are related to
the conformal curvatures. For example, the torsion tensors are related by
\begin{align}
T^a = \poin T^a, \quad
T^\alpha{}_i = \poin T^\alpha{}_i + i E^b \wedge F_{\dgamma i} \,\bsigma_b^{\dgamma \alpha}, \quad
T_{\dalpha}{}^i = \poin T_\dalpha{}^i + i E_b \wedge F^{\gamma i} \,\sigma^b_{\gamma \dalpha}.
\end{align}
The difference in the terms is so simple that it is clear to see
that the relations
\begin{gather}
T_{\balpha \bbeta}{}^C =  T^{\dbalpha \dbbeta C} =  T_{\balpha}{}^{\dbbeta \bgamma} = 
       T_{\balpha}{}^{\dbbeta}{}_{\dbgamma} = 0 \eol
T_\alpha{}^i{}^\dbeta{}_j{}^c = 2i \delta^i_j (\sigma^c)_\alpha{}^\dbeta \eol
T_{A b}{}^c = 0, \quad
T_{\alpha \dalpha\, \beta}{}^j{}^\gamma{}_k = \frac{1}{2} \delta_\alpha{}^\gamma \,T_{\phi \dalpha\, \beta}{}^j{}^\phi{}_k 
\end{gather}
obeyed by the conformal torsion tensor are \emph{also} obeyed by the
Poincar\'e torsion tensors. These relations are precisely the torsion
tensor constraints \eqref{eq_Tconstraints} which together with the
choice of structure group uniquely determine Howe's formulation of
conformal supergravity. Our structure \emph{must} therefore
correspond to Howe's solution when degauged. We will now show
this explicitly by solving the constraints obeyed by
the connections $F_A{}^B$.

\subsection{The conformal origin of $\N=2$ auxiliary superfields}
In the gauge where $B_M=0$, the conformal dilatation curvature is
given by
\begin{align}
R(\D)_{BA} = + 2 F_{BA}(-)^a - 2 F_{AB} (-)^{b+ab}.
\end{align}
The purely chiral and antichiral spinor curvatures vanish, giving
\begin{subequations}
\begin{align}
R(\D)_{\bbeta \balpha} = 0 &\implies F_{{\bbeta \balpha}} = -F_{\balpha \bbeta} \\
R(\D)^{\dbbeta \dbalpha} = 0 &\implies F^{{\dbbeta \dbalpha}} = -F^{\dbalpha \dbbeta}.
\end{align}
\end{subequations}
These imply the Lorentz and isospin decompositions
\begin{subequations}
\begin{align}
F_{\beta}{}^j{}_\alpha{}^i &= \frac{1}{2} \eps_{\beta \alpha} S^{ji} - \frac{1}{2} \eps^{ji} Y_{\beta \alpha} \\
F^{\dbeta}{}_j{}^\dalpha{}_{i}  &= \frac{1}{2} \eps^{\dbeta \dalpha} \bar S_{ji}
     + \frac{1}{2} \eps_{ji} \bar Y^{\dbeta \dalpha}.
\end{align}
\end{subequations}
The mixed spinor curvature also vanishes, giving
\begin{align}
R(\D)_{\bbeta}{}^\dbalpha = 0 &\implies F_{\bbeta}{}^\dbalpha = -F^\dbalpha{}_\bbeta.
\end{align}
This implies the isospin decompositions
\begin{subequations}
\begin{align}
F_{\alpha}{}^i{}_{\dalpha j}  = +\delta^i_j G_{\alpha \dalpha} + i G_{\alpha\dalpha}{}^i{}_j \\
F^{\dalpha}{}_i{}^{\alpha j}  = -\delta_i^j G^{\dalpha \alpha} - i G^{\dalpha\alpha}{}^j{}_i
\end{align}
\end{subequations}
where $G_c$ and $G_c{}^{ij}$ are real isosinglet and isotriplet vectors, respectively.

In identifying these superfields as elements of the special conformal connections
of the gauge-fixed geometry, we can give an alternative explanation of their super-Weyl
transformation properties. Because Howe's structure is given by the
choice $B_M = 0$, any dilatation must be accompanied by a special conformal
transformation which restores this gauge choice. We find
\begin{align}
0 = \delta_{SW}(\Lambda) B_M &= \delta_D(\Lambda) B_M + \delta_K(\eps^A) B_M 
\end{align}
for the choice
\begin{align}
\eps^a = \frac{1}{2} \CD^a \Lambda, \qquad
\eps^{\alpha j} = -\frac{1}{2} \CD^{\alpha j} \Lambda, \qquad
\eps_{\dalpha j} = -\frac{1}{2} \BCD_{\dalpha j} \Lambda.
\end{align}
The $\delta_K$ in the above expression generates the inhomogeneous
part of the super-Weyl transformation. Using
$\delta_K F_M{}^{\beta \,j}= \CD_M \eps^{\beta j} - i E_{M \dbeta}{}^j \eps^{\dbeta \beta}$
and its conjugate, we find
\begin{subequations}
\begin{align}
\delta_K S^{ij} &= \delta_K F^{\alpha i}{}_\alpha{}^j
     = \CD^{\alpha i} \eps_\alpha{}^j = -\frac{1}{2} \CD^{ij} \Lambda \\
\delta_K Y_{\alpha \beta} &= \delta_K F_\beta{}^k{}_{\alpha\, k}
     = -\frac{1}{2} \CD_{\beta \alpha} \Lambda \\
\delta_K G_{\alpha \dalpha} &= \frac{1}{2} \delta_K F_{\alpha}{}^k{}_{\dalpha\, k}
     = \frac{1}{2} \CD_\alpha{}^k \eps_{\dalpha k} - i \eps_{\alpha \dalpha}
     = - \frac{1}{8} [\CD_\alpha{}^k, \BCD_{\dalpha k}] \Lambda \\
\delta_K G_{\alpha \dalpha}{}^{ij} &= -i \delta_K F_{\alpha}{}^{(i}{}_\dalpha{}^{j)}
     = \frac{i}{4} [\CD_\alpha{}^{(i}, \CD_\dalpha{}^{j)}] \Lambda
\end{align}
\end{subequations}
which are precisely the inhomogeneous parts of the transformation laws 
for these objects as given in \eqref{eq_HoweSW}.

\subsection{The conformal origin of constraints}
The superfields of Howe's superspace formulation are constrained by a set of
dimension-3/2 Bianchi identities. It turns out that these relations are encoded
in the structure of conformal superspace in the constraints on the dimension-3/2
dilatation and conformal curvatures.

We begin by considering the dimension-3/2 dilatation curvature. In the gauge
where $B_M=0$, we have
\begin{align}
R(\D)_{\beta}{}^j{}_{\alpha \dalpha} =
     -\frac{i}{2} \eps_{\beta \alpha} \bar\nabla_\dphi{}^j \bar W^\dphi{}_\dalpha
     = 2 F_{\beta}{}^j{}_{\alpha \dalpha} + 2 F_{\alpha \dalpha\,}{}_\beta{}^j
\end{align}
and similarly for the conjugate expression, yielding
\begin{subequations}
\begin{align}
F_{\alpha \dalpha\,}{}_{\beta}{}^j + F_{\beta}{}^j{}_{\,\alpha \dalpha} = 
     - \frac{i}{4} \eps_{\beta \alpha} \bar\nabla_\dphi{}^j \bar W^\dphi{}_\dalpha \\
F_{\alpha \dalpha\,}{}_{\dbeta j} + F_{\dbeta j \,\alpha \dalpha} =
     - \frac{i}{4} \eps_{\dbeta \dalpha} \nabla^\phi{}_j W_{\phi \alpha}.
\end{align}
\end{subequations}

Next, we need to consider the dimension-3/2 components of the special
conformal curvatures. In general, we have\footnote{The gradings
have been suppressed. In the first term of each expression, $[CB]$
means $CB - BC (-)^{bc}$ whereas in each of the third terms, $[CB]$ means
$(-)^b CB - BC (-)^{bc + c}$.}
\begin{subequations}
\begin{align}
R(K)_{CB}{}^a &= \CD_{[C} F_{B]}{}^a + \poin T_{CB}{}^D F_D{}^a
     + 2i F_{[C}{}^{\alpha j} F_{B]}{}^\dalpha{}_j \,\sigma^a_{\alpha \dalpha} \\
R(K)_{CB}{}^{\alpha i} &= \CD_{[C} F_{B]}{}^{\alpha i}
     + \poin T_{CB}{}^D F_D{}^{\alpha i}
     - i \delta_{[C\, \dalpha}{}^i F_{B]}{}^a \,\bsigma_a^{\dalpha \alpha} \\
R(K)_{CB \,\dalpha i} &= \CD_{[C} F_{B] \dalpha i}
     + \poin T_{CB}{}^D F_{D \dalpha i}
     - i \delta_{[C}{}^{\alpha}{}_i F_{B] a} \,\sigma^a_{\alpha \dalpha}
\end{align}
\end{subequations}
where $\poin T_{CB}{}^D$ is the Poincar\'e torsion tensor.
There are three independent dimension-3/2 constraints among the special conformal curvatures.
The simplest is
\begin{align}
0 = R_{\bbeta \balpha}{}^{\gamma k} \equiv
     \CD_\bbeta F_\balpha{}^{\gamma k}
     + \CD_\balpha F_\bbeta{}^{\gamma k}.
\end{align}
Plugging in the form of $F_{\beta}{}^j{}_{\gamma}{}^k$, we find
\begin{gather}
\CD_\beta{}^{(k} S^{ij)} = 0, \quad
\CD_{(\gamma}{}^j Y_{\beta\alpha)} = 0, \quad
\CD_{\alpha j} S^{ji} = \CD^{\beta i} Y_{\beta \alpha}.
\end{gather}

The second constraint is
\begin{align}
2 \eps_{\beta \alpha} \eps^{ji} \cW(S)_{\dgamma k} =
     R_{\beta}{}^j{}_\alpha{}^i{}_{\, \dgamma k} \equiv
     \CD_\bbeta F_{\balpha \,\dgamma k}
     + \CD_\balpha F_{\bbeta \,\dgamma k}
     + i \delta^j_k F_\alpha{}^i{}_{\beta \dalpha}
     + i \delta^i_k F_\beta{}^j{}_{\alpha \dalpha}.
\end{align}
Inserting the form of $F_{\balpha\, \dgamma k}$ yields the
solution
\begin{align}
i F_{\beta}{}^j{}_{\,\alpha \dalpha}
     = -\CD_\beta{}^j G_{\alpha \dalpha}
     - \frac{i}{3} \CD_{(\beta}{}^k G_{\alpha) \dalpha}{}^j{}_k
     + \frac{i}{2} \eps_{\beta \alpha} \CD^{\phi k} G_{\phi \dalpha}{}^j{}_k
     + 2 \eps_{\beta \alpha} \bar \cW(S)_\dalpha{}^j
\end{align}
along with the additional constraint
\begin{align}
\CD_{(\beta}{}^{(k} G_{\alpha) \dalpha}{}^{ij)} = 0.
\end{align}

Finally, the constraint for $R(S)_{\bbeta}{}^\dbalpha{}^{\gamma k}$ gives
\begin{align}
0 = R(S)_\beta{}^j{}_{\dalpha i\, \gamma}{}^k
     = \CD_\beta{}^j F_{\dalpha i\, \gamma}{}^k
     + \BCD_{\dalpha i} F_\beta{}^j{}_\gamma{}^k
     + 2i \delta_i^j F_{\beta \dalpha\, \gamma}{}^k
     + i \delta_i^k F_\beta{}^j{}_{\, \gamma \dalpha}.
\end{align}
Plugging in the form for each of these superfields leads to
\begin{gather}
\BCD_\dalpha{}^{(i} S^{jk)} = i \CD^{\phi (i} G_{\phi \dalpha}{}^{jk)} \\
\CD_\beta{}^j G_{\alpha \dalpha}
     = -\frac{1}{4} \BCD_\dalpha{}^j Y_{\beta \alpha}
     + \frac{1}{12} \eps_{\beta \alpha} \BCD_{\dalpha k} S^{kj}
     + \frac{i}{3} \eps_{\beta \alpha} \CD^{\phi k} G_{\phi \dalpha}{}^j{}_k 
     + \frac{1}{4} \eps_{\beta \alpha} \BCD_\dphi{}^j \bar W^{\dphi}{}_\dalpha.
\end{gather}
This reproduces all of the dimension-3/2 constraints in Howe's
superspace geometry \cite{Howe, KLRT_M_Howe}.

In the interest of fully expressing the degauged parameters,
we should also mention how to solve for $F_b{}^a$, which is a dimension two
superfield. It is easily found in the constraint for
$R(K)_{\bgamma}{}^{\dbbeta\, a}$
\begin{align}
0 = R(K)_{\gamma}{}^k{}_{\dbeta\,j\, \alpha \dalpha}
     = \CD_\gamma{}^k F_{\dbeta\, j\, \alpha \dalpha}
     + \BCD_{\dbeta j} F_\gamma{}^k{}_{\alpha \dalpha}
     + 2i \delta^k_j F_{\gamma \dbeta\, \alpha \dalpha}
     + 4i F_{\gamma}{}^{k}{}_\alpha{}^{l} F_{\dbeta\, j}{}_{\dalpha\, l}
     + 4i F_{\dbeta\, j}{}_\alpha{}^{l} F_{\gamma}{}^k{}_{\dalpha\, l}
\end{align}
in terms of quantities already defined.

\subsection{Reproducing degauged curvatures}
It is a straightforward task to reproduce the curvatures of the
$\rm SL(2,\mathbb C) \times \rm U(2)_R$ geometry via this same
degauging process. The most straightforward way is to take the
curvature two-forms in the conformal geometry and write them as the
degauged curvatures. For example, suppose we want to calculate the
components of the isospin curvature tensor $R(I)_{\bbeta \balpha\, kl}$.
Using its explicit form in terms of the superconnections \eqref{eq_RI2form},
we may identify
\begin{align}
R(I)_{\beta}{}^j{}_\alpha{}^i{}_{kl} &=
     \hat R(I)_{\beta}{}^j{}_\alpha{}^i{}_{kl}
     + 4 \delta^j_{(k} F_{\alpha}{}^i{}_{\beta l)}
     + 4 \delta^i_{(k} F_{\beta}{}^j{}_{\alpha l)}.
\end{align}
Because this curvature is constrained to be zero, we may identify the
degauged curvature
\begin{align}
\hat R(I)_{\beta}{}^j{}_\alpha{}^i{}_{kl} =
     - 4 \delta^j_{(k} F_{\alpha}{}^i{}_{\beta l)}
     - 4 \delta^i_{(k} F_{\beta}{}^j{}_{\alpha l)}
     = -4 \delta^j_{(k} \delta^i_{l)} Y_{\alpha \beta}.
\end{align}

An equivalent (but more efficient) way of proceeding is to calculate
the desired commutators directly. For example, to calculate
$\{\CD_\alpha{}^i, \CD_\beta{}^j\}$, one may consider a conformally
primary field $\Psi$ lying in some arbitrary representation of the
rest of the superconformal algebra. Then observing that
\begin{align}
\nabla_\balpha = \CD_\balpha - F_\balpha{}^C K_C
\end{align}
we may calculate
\begin{align}
\nabla_\balpha \nabla_\bbeta \Psi
     &= \CD_\balpha \nabla_\bbeta \Psi - F_\balpha{}^C [K_C, \nabla_\bbeta] \Psi \eol
     &= \CD_\balpha \CD_\bbeta \Psi - F_\balpha{}^C [K_C, \nabla_\bbeta] \Psi.
\end{align}
Symmetrizing the indices and reordering the expression, we are led to
\begin{align}
\{\CD_\balpha, \CD_\bbeta\} \Psi = \{\nabla_\balpha, \nabla_\bbeta\} \Psi
     + F_\balpha{}^C [K_C, \nabla_\bbeta] \Psi
     + F_\bbeta{}^C [K_C, \nabla_\balpha] \Psi.
\end{align}
Since the only special conformal generator in the commutator
which will yield a non-vanishing result is $S$, we have
\begin{align}
\{\CD_\alpha{}^i, \CD_\beta{}^j\} \Psi &= -\eps_{\alpha \beta} \eps^{ij} \bar W_{\dalpha \dbeta} M^{\dbeta \dalpha}  \Psi
     + F_\alpha{}^i{}^{\gamma k} \{S_{\gamma k}, \nabla_\beta{}^j\} \Psi
     + F_\beta{}^j{}^{\gamma k} \{S_{\gamma k}, \nabla_\alpha{}^i\} \Psi.
\end{align}
Evaluating the commutators and inserting the forms for $F$ gives
\begin{align}
\{\CD_\alpha{}^i, \CD_\beta{}^j\} &= -\eps_{\alpha \beta} \eps^{ij} \bar W_{\dalpha \dbeta} M^{\dbeta \dalpha} 
     + 2 S^{ij} M_{\alpha \beta}
     + \eps^{ij} \eps_{\alpha \beta} Y^{\delta \phi} M_{\phi \delta}
     \eol & \quad
     + 2 \eps_{\alpha \beta} \eps^{ij} S^k{}_l I^l{}_k
     - 4 Y_{\alpha \beta} I^{ij}.
\end{align}
All of the commutators may be evaluated in either way.

\subsection{Action principle for $\rm SL(2,\mathbb C) \times \rm U(2)_R$ supergravity}\label{Howe_action}
In the previous section we have given the form for the component
evaluation of a chiral action as well as the rule for transforming
a full superspace action into a chiral one. These are well-known
procedures in $\rm U(2)_R$ supergravity; since we may degauge our
formulae to regular $\rm U(2)_R$ supergravity, it is a straightforward
exercise to verify that we reproduce these results.

Recall that a full conformal superspace action may be converted
to a chiral action in a very simple way:
\begin{align}
\int d^{12}z\, E\, \Lag = \int d^{8}\chz\, \chE\, \bar\nabla^4 \Lag, \qquad
\bar\nabla^4\equiv \frac{1}{48} \bar\nabla^{ij}\bar\nabla_{ij}.
\end{align}
Because of the simplicity of the conformal superspace curvatures,
it is simple to show that we could just as easily have written
\begin{align}
\bar\nabla^4 \Lag = \frac{1}{96} \bar\nabla^{ij}\bar\nabla_{ij} \Lag 
     - \frac{1}{96} \bar\nabla^{\dalpha \dbeta}\bar\nabla_{\dalpha \dbeta} \Lag
\end{align}
since the two terms are equivalent in conformal supergravity when
$\Lag$ is real with vanishing conformal weight.
We would like to degauge both of them. This is most easily
done working from the outside in. Keeping in mind that $\Lag$
has vanishing conformal and $U(1)_R$ weights, we find for
the first term
\begin{align*}
\bar\nabla^{ij} \bar\nabla_{ij} \Lag
     &= \BCD_\dalpha{}^i \bar\nabla^{\dalpha j} \bar\nabla_{ij} \Lag
     + 6 \bar S_{ij} \bar\nabla^{ij} \Lag + 6 \bar Y_{\dalpha \dbeta} \bar\nabla^{\dbeta \dalpha} \Lag \eol
     &= \BCD_\dalpha{}^i (\BCD^{\dalpha\,j} \bar\nabla_{ij} \Lag
               - 2 \bar S_{ij} \bar\nabla^{\dalpha j} \Lag + 6 \bar Y^{\dalpha \dbeta} \bar\nabla_\dbeta{}_i \Lag)
     + 6 \bar S_{ij} \bar\nabla^{ij} \Lag + 6 \bar Y_{\dalpha \dbeta} \bar\nabla^{\dbeta \dalpha} \Lag.
\end{align*}
Observing that $\bar\nabla^\dalpha{}_i \Lag = \CD^\dalpha{}_i\Lag$
and $\bar\nabla_{ij} \Lag = \BCD_{ij} \Lag$, we find
\begin{align*}
\bar\nabla^{ij} \bar\nabla_{ij} \Lag
     &= \BCD_{ij} \BCD^{ij} \Lag + 4 \bar S_{ij} \BCD^{ij} \Lag
     + 12 \bar Y_{\dalpha \dbeta} \BCD^{\dalpha \dbeta} \Lag
     - 2 \CD_\dalpha{}^i \bar S_{ij} \BCD^{\dalpha j} \Lag
     + 6 \BCD_\dalpha{}^j \bar Y^{\dalpha \dbeta} \BCD_{\dbeta j} \Lag.
\end{align*}
Similarly we may calculate
\begin{align*}
\bar\nabla_{\dalpha\dbeta} \bar\nabla^{\dalpha \dbeta} \Lag
     &= \BCD_{\dalpha \dbeta} \BCD^{\dalpha \dbeta} \Lag
     - 4 \bar Y_{\dalpha \dbeta} \BCD^{\dbeta \dalpha} \Lag
     - 12 \bar S_{ij} \BCD^{ij} \Lag
     - 6 \BCD_\dalpha{}^j \bar S_{jk} \BCD^{\dalpha k} \Lag
     + 2 \BCD_\dalpha{}^j \bar Y^{\dalpha \dbeta} \BCD_{\dbeta j} \Lag.
\end{align*}
Putting these together, we find
\begin{align}\label{eq_ChiralProjectionOperator}
\bar\nabla^4 \Lag = \frac{1}{96} (\BCD_{ij} \BCD^{ij} - \BCD_{\dalpha \dbeta} \BCD^{\dalpha \dbeta}) \Lag
     + \frac{1}{6} \bar S_{ij} \BCD^{ij} \Lag
     + \frac{1}{6} \bar Y_{\dalpha \dbeta} \BCD^{\dalpha \dbeta} \Lag
\end{align}
which is precisely the $\N=2$ chiral projection operator as given
in \cite{Muller}.

By a very nearly identical line of attack, one may degauge the
component action \eqref{eq_Lagc} to reproduce the component action
originally calculated by M\"uller \cite{Muller2}. We leave this as an
exercise to the interested reader. (See \cite{Gates:2009xt}
for an alternative calculation of this same component Lagrangian
in notation more closely related to our own.)

\section{Future directions and outlook}
We have presented a single framework for $\N=2$ supergravity
which is simultaneously manifestly superconformal and manifestly
supersymmetric. It reduces in components to the manifestly
superconformal framework of de Wit et al.
\cite{sct_chiral, sct_gaugematter, sct_imptensor, sct_lagrangians, sct_rules, sct_structure}
and can be degauged in superspace to the manifestly supersymmetric
framework constructed by Howe \cite{Howe}.
At the same time, it sheds light on certain curious features of both
these formulations.

It is a very pleasant surprise that the basic principles of our
earlier work on $\N=1$ conformal superspace \cite{Butter:2009cp}
can so easily be applied to $\N=2$. This raises two very interesting questions.
First, can we extend this formulation somehow to $\N=4$?
Second, can we extend this to different dimensions?
One very nice feature of this superconformal approach is that it can
so readily reproduce the geometric structure of Poincar\'e supergravity.
For the cases of $\N=1,2$ in four dimensions which we have explored,
this reproduction isn't terribly impressive since it tells us nothing
new, merely reconfirming old results in a new light. However, for other dimensions
where the superspace structure of supergravity remains
something of a mystery, it is a distinct possibility that superconformal procedures
may offer up superspace secrets more quickly and more readily than
conventional approaches.

Even within the context of $\N=2$ theories in four dimensions, there
remains a great deal to understand. There has been a recent flurry of
activity in the supergravity structure of projective superspace, in 
particular its superconformal
properties \cite{Kuzenko:2007qy, KLRT_M_Grimm, KLRT_M_Howe}. It seems
very likely that the manifestly superconformal framework constructed here
may offer insight. We have already begun to investigate
projective superspace with a superconformal structure group and have
found some promising early results. For example, the constraints
\eqref{eq_gaugeconstraints} can be naturally interpreted as integrability
conditions for the existence of covariantly analytic superfields:
\begin{align}
\{\nabla_\alpha^+, \nabla_\beta^+\} = \{\nabla_\alpha^+, \bar\nabla_\dbeta^+\} = 0, \qquad
\nabla_\alpha^+ \equiv v_i \nabla_\alpha{}^i, \qquad
\bar\nabla_\dalpha^+ \equiv v_i \bar\nabla_\dalpha{}^i.
\end{align}
Similarly, it becomes natural to interpret $I^i{}_j$ of the superconformal
algebra as a differential operator on the isotwistors $v^i$.

It would also be very interesting to understand how this
superconformal approach works in harmonic superspace, where superconformal properties
are less straightforwardly realized than in projective superspace.\footnote{See for
example the discussion in Chapter 9 of \cite{GIOS}.} \\

\noindent
{\bf Acknowledgements:}\\
The author would like to thank Sergei Kuzenko for helpful discussions.
This work is supported by the Australian Research Council (grant No. DP1096372)
and by a UWA Research Development Award.

\newpage
\appendix
\section{Solution to the $\N=2$ Bianchi identities}
The constraints chosen for $\N=2$ conformal supergravity in superspace
are a subset of a set of constraints identical in form to super Yang-Mills in
flat superspace \cite{Grimm:1977xp}. The constraints amount to
the specification of the spinor derivative anticommutators:
\begin{gather}
\{\nabla_\alpha{}^i, \nabla_\beta{}^j\} = -2 \eps_{\alpha \beta} \eps^{ij} \bar\cW, \quad
\{\bar\nabla^\dalpha{}_i, \bar\nabla^\dbeta{}_j\} = +2 \eps^{\dalpha \dbeta} \eps_{ij} \cW \eol
\{\nabla_\alpha{}^i, \bar\nabla_{\dbeta j}\} = -2i \delta^i_j \nabla_{\alpha \dbeta} \label{eq_BianchiConstraint}
\end{gather}
where $\cW$ and $\bar \cW$ are (for the moment) arbitrary operators related
by complex conjugation.
In order for this structure to be consistent, we must examine all the Bianchi
identities:
\begin{align*}
0 = \sum_{[ABC]} [\nabla_A, [\nabla_B, \nabla_C]]
\end{align*}
where the sum is over graded cyclic permutations of the indices.

There are two independent dimension-3/2 Bianchi identities. The first is
\begin{align*}
0 &= [\nabla_\alpha{}^i, \{\nabla_\beta{}^j, \nabla_\gamma{}^k\}] +
     [\nabla_\beta{}^j, \{\nabla_\gamma{}^k, \nabla_\alpha{}^i\}] +
     [\nabla_\gamma{}^k, \{\nabla_\alpha{}^i, \nabla_\beta{}^j\}] \eol
     &= - 2\eps_{\beta \gamma} \eps^{jk} [\nabla_\alpha{}^i, \bar\cW]
     - 2 \eps_{\gamma \alpha} \eps^{ki} [\nabla_\beta{}^j, \bar\cW]
     - 2 \eps_{\alpha \beta} \eps^{ij} [\nabla_\gamma{}^k, \bar \cW].
\end{align*}
The constraint is satisfied provided
\begin{align}
0 = [\nabla_\alpha{}^i, \bar \cW].
\end{align}
The second dimension-3/2 Bianchi identity is
\begin{align*}
0 &= [\nabla_\alpha{}^i, \{\nabla_\beta{}^j, \bar\nabla_{\dgamma k} \}] +
     [\nabla_\beta{}^j, \{\bar\nabla_{\dgamma k}, \nabla_\alpha{}^i\}] +
     [\bar\nabla_{\dgamma k}, \{\nabla_\alpha{}^i, \nabla_\beta{}^j\}] \eol
     &= -2i \delta^j_k [\nabla_\alpha{}^i, \nabla_{\beta \dgamma}]
     -2i \delta^i_k [\nabla_\beta{}^j, \nabla_{\alpha \dgamma}]
     - 2 \eps_{\alpha \beta} \eps^{ij} [\bar\nabla_{\dgamma k}, \bar\cW].
\end{align*}
Symmetrizing $\alpha$ and $\beta$, one may easily see that
\begin{align}
[\nabla_\alpha{}^i, \nabla_{\beta \dbeta}] = -2 \eps_{\alpha \beta} \bar\cW_\dbeta{}^i
\end{align}
for some operator $\bar\cW_\dbeta{}^i$. Reinserting this result into the original
expression then yields
\begin{align}
\bar \cW_\dbeta{}^i = -\frac{i}{2} [\bar\nabla_{\dbeta}{}^i, \bar\cW].
\end{align}
The other dimension-3/2 Bianchi identities are related to these by complex
conjugation,
\begin{align}
[\bar\nabla_{\dalpha i}, \nabla_{\beta \dbeta}] = -2 \eps_{\alpha \beta} \cW_{\beta i}, \quad
\cW_{\beta i} = -\frac{i}{2} [\nabla_{\beta i}, \cW].
\end{align}

There are two independent dimension-2 Bianchi identities. The first is
\begin{align*}
0 = \{\nabla_\alpha{}^i, [\nabla_\beta{}^j, \nabla_{\gamma \dgamma} ]\}
     + [\nabla_{\gamma \dgamma}, \{\nabla_\alpha{}^i, \nabla_\beta{}^j\}]
     - \{\nabla_\beta{}^j, [\nabla_{\gamma \dgamma}, \nabla_\alpha{}^i]\}
\end{align*}
and it is satisfied automatically given the prior constraints.
The second is
\begin{align*}
0 &= \{\nabla_\alpha{}^i, [\bar\nabla_{\dbeta j}, \nabla_{\gamma \dgamma} ]\}
     + [\nabla_{\gamma \dgamma}, \{\nabla_\alpha{}^i, \bar\nabla_{\dbeta j}\}]
     - \{\bar\nabla_{\dbeta j}, [\nabla_{\gamma \dgamma}, \nabla_\alpha{}^i]\} \eol
     &= i \eps_{\dbeta \dgamma} \{\nabla_\alpha{}^i, [\nabla_{\gamma j}, W]\}
     -2i \delta^i_j [\nabla_{\gamma \dgamma}, \nabla_{\alpha \dbeta}]
     + i \eps_{\alpha \gamma} \{\bar\nabla_{\dbeta j}, [\bar \nabla_{\dgamma}{}^i, \bar\cW]\}
\end{align*}
By contracting all the spinor indices, one may show that
\begin{align}
\{\nabla^{\phi i}, [\nabla_{\phi j}, \cW]\} = \{\bar\nabla_{\dphi j}, [\bar\nabla^{\dphi i} , \bar\cW]\}.
\end{align}
Making use of this result in the prior expression, one finds
\begin{align}
[\nabla_{\alpha \dalpha}, \nabla_{\beta \dbeta}]
     = - \frac{1}{4} \eps_{\dalpha \dbeta} \{\nabla_{(\alpha}{}^k, [\nabla_{\beta) k}, \cW]\}
     + \frac{1}{4} \eps_{\alpha \beta} \{\bar\nabla_{(\dalpha}{}^k [\bar\nabla_{\dbeta) k}, \bar\cW]\}.
\end{align}

The remaining dimension-5/2 and dimension-3 Bianchi identities lead to no
new results.

\section{Curvatures of conformal superspace}\label{app_Curvatures}
In this section, we summarize the results for the conformal curvatures.
The conformal covariant derivative is given by
\begin{align}
\nabla_A = E_A{}^M
     \left(\partial_M - \frac{1}{2} \Omega_M{}^{cb} M_{bc}
     - \Phi_M{}^i{}_j I^j{}_i
     - A_M \A - B_M \D 
     - F_M{}^B K_B \right)
\end{align}
and its curvatures are
\begin{align}
[\nabla_A, \nabla_B] = -T_{AB}{}^C \nabla_C
     - \frac{1}{2} R_{AB}{}^{cd} M_{dc}
     - R_{AB}{}^i{}_j I^j{}_i
     - R(\A)_{AB} \A
     - R(\D)_{AB} \D
     - R_{AB}{}^C K_C.
\end{align}
We summarize the results for each curvature.

\subsection{Torsion}
The torsion two-forms are defined in terms of the gauge connections as
\begin{align}
T^a &= dE^a + E^b \wedge \Omega_b{}^a + E^a \wedge B \\
T^\alpha{}_i &= dE^\alpha{}_i + \frac{1}{2} E^\alpha{}_i \wedge B
     - i E^\alpha{}_i \wedge A + E^\beta{}_i \wedge \Omega_\beta{}^\alpha
     + E^\alpha{}_j \wedge \Phi^j{}_i
     + i  E^b \wedge F_{\dgamma i} \,\bsigma_b^{\dgamma \alpha} \\
T_\dalpha{}^i &= dE_\dalpha^i + \frac{1}{2} E_\dalpha{}^i \wedge B
     + i E_\dalpha{}^i \wedge A + E_\dbeta{}^i \wedge \Omega^\dbeta{}_\dalpha
     - E_\dalpha{}^j \wedge \Phi_j{}^i
     + i  E_b \wedge F^{\gamma i} \,\sigma^b_{\gamma \dalpha}
\end{align}
The components $T_{\ul \bgamma \ul \bbeta}{}^A$ are given by
\begin{gather}
T_{\ul\bgamma \ul\bbeta}{}^{\ul\balpha} = 0 \\
T_{\bgamma \bbeta}{}^a = T^{\dbgamma \dbbeta a} = 0, \quad
T_{\gamma}{}^i{}^{\dbeta}{}_j{}^a = -2i \,\delta^i_j \,{(\sigma^a \eps)_{\gamma}}^{\dbeta}
\end{gather}
The components
$T_{\ul \bgamma b}{}^A = -\frac{1}{2} \bsigma_b^{\dbeta \beta} T_{\ul \bgamma (\beta \dbeta)}{}^A$
are given by
\begin{gather}
?T_{\ul\bgamma}_b^a? = 0 \\
?T_{\bgamma}_b^\balpha? = ?T^{\dbgamma}_b_\dbalpha? = 0 \\
?T_\bgamma_b_\dbalpha? \ \  \leadsto \ \
T_\gamma{}^j{}_{\,\beta \dbeta\,}{}_\dalpha{}^i = i \eps_{\gamma \beta} \eps^{ji} \bar W_{\dbeta \dalpha} \\
?T^\dbgamma_b^\balpha? \ \  \leadsto \ \
T_{\dgamma}{}^j{}_{\,\beta \dbeta\,}{}_{\alpha}{}^i = -i \eps_{\dgamma \dbeta} \eps^{ji} W_{\beta \alpha}
\end{gather}
The components $T_{cb}{}^A = \frac{1}{4} \bsigma_c^{\dgamma \gamma} \bsigma_b^{\dbeta \beta}
     T_{\gamma \dgamma \,\beta \dbeta\,}{}^A$ are given by
\begin{gather}
T_{cb}{}^a = 0 \\
{T_{cb}}^\balpha \ \  \leadsto \ \
     T_{\gamma \dgamma \,\beta \dbeta\,}{}_\alpha{}^i =
          \frac{1}{2} \eps_{\dgamma \dbeta} \nabla_\alpha{}^i W_{\gamma \beta} \\
?T_c_b_\dbalpha? \ \  \leadsto \ \
     T_{\gamma \dgamma \,\beta \dbeta\,}{}_\dalpha{}^i =
          \frac{1}{2} \eps_{\gamma \beta} \bar\nabla_\dalpha{}^i \bar W_{\dgamma \dbeta}
\end{gather}

\subsection{Lorentz curvature}
The conformal Lorentz curvature two-form is
\begin{align}
R^{ba} = d\Omega^{ba} + \Omega^{bc} \wedge \Omega_c{}^a - 2 E^{[b} \wedge F^{a]}
     - 4 E^{\beta}{}_i \wedge F^{\alpha i} \,(\sigma^{ba} \eps)_{\alpha\beta}
     - 4 E_{\dbeta}{}^i \wedge F_{\dalpha i} \,(\bsigma^{ba} \eps)^{\dalpha\dbeta}
\end{align}
and may be canonically decomposed
\begin{align}
{R_{DC}}^{ba} \ \ \leadsto \ \ R_{DC \,\beta\dbeta\, \alpha \dalpha} =
2 \eps_{\dbeta \dalpha} R_{DC \beta \alpha}
- 2 \eps_{\beta \alpha} R_{DC \dbeta \dalpha}.
\end{align}
It is simplest to express the curvature results in terms of these components.
We group them by dimension.

\begin{itemize}
\item {Dimension 1}
\begin{gather}
R_{\bdelta \bgamma \, \beta \alpha} = 0, \quad
R_{\bdelta \bgamma \, \dbeta \dalpha} = 2 C_{\bdelta \bgamma} \bar W_{\dbeta \dalpha} \\
R_{\dbdelta \dbgamma \, \dbeta \dalpha} = 0, \quad
R^{\dbdelta \dbgamma}{}_{\beta \alpha} = -2 C^{\dbdelta \dbgamma} W_{\beta \alpha} \\
R_{\bdelta}{}^{\dbgamma}{}_{\beta \alpha} = R_{\bdelta}{}^{\dbgamma}{}_{\dbeta \dalpha} = 0
\end{gather}
\item {Dimension 3/2}
\begin{gather}
R_{\delta}{}^j{}_{\,\gamma \dgamma\,}{}_{\beta \alpha} = 0, \quad
R_{\delta}{}^j{}_{\,\gamma \dgamma\,}{}_{\dbeta \dalpha} =
     - \frac{i}{2} \eps_{\delta \gamma} \bar\nabla_\dbeta{}^j \bar W_{\dalpha \dgamma}
     - \frac{i}{2} \eps_{\delta \gamma} \bar\nabla_\dalpha{}^j \bar W_{\dbeta \dgamma} \\
R_{\ddelta}{}^j{}_{\,\gamma \dgamma\,}{}_{\dbeta \dalpha} = 0, \quad
R_{\ddelta}{}^j{}_{\,\gamma \dgamma\,}{}_{\beta \alpha} =
     - \frac{i}{2} \eps_{\ddelta \dgamma} \nabla_\beta{}^j W_{\alpha \gamma}
     - \frac{i}{2} \eps_{\ddelta \dgamma} \nabla_\alpha{}^j W_{\beta \gamma}
\end{gather}
\item{Dimension 2}
\begin{align}
R_{\delta \ddelta\,\gamma \dgamma\,\beta \alpha} &=
     + \frac{1}{4} \eps_{\ddelta \dgamma} \nabla_{\beta \alpha} W_{\delta \gamma}
     - \frac{1}{8} \eps_{\ddelta \dgamma}
          (\eps_{\delta \beta} \eps_{\gamma \alpha} + \eps_{\delta \alpha} \eps_{\gamma \beta})
          \nabla_{\phi \mu} W^{\mu \phi}
     + \eps_{\delta \gamma} \bar W_{\ddelta \dgamma} W_{\beta \alpha} \\
R_{\delta \ddelta \,\gamma \dgamma\, \dbeta \dalpha} &=
     - \frac{1}{4} \eps_{\delta \gamma} \bar \nabla_{\dbeta \dalpha} \bar W_{\ddelta \dgamma}
     + \frac{1}{8} \eps_{\delta \gamma}
          (\eps_{\ddelta \dbeta} \eps_{\dgamma \dalpha} + \eps_{\ddelta \dalpha} \eps_{\dgamma \dbeta})
          \bar \nabla_{\dphi \dmu} \bar W^{\dmu \dphi}
     - \eps_{\ddelta \dgamma} W_{\delta \gamma} \bar W_{\dbeta \dalpha}
\end{align}
\end{itemize}

\subsection{Dilatation and $\rm U(1)_R$ curvatures}
The conformal field strengths for scalings and chiral rotations are
\begin{align}
R(\D) &= dB + 2 E^a \wedge F_a - 2 E^\balpha \wedge F_\balpha - 2 E_\dbalpha \wedge F^\dbalpha \\
R(\A) &= dA + i E^\balpha \wedge F_\balpha - i E_\dbalpha \wedge F^\dbalpha 
\end{align}
We group the results by dimension.
\begin{itemize}
\item {Dimension 1}
\begin{gather}
R(\D)_{\ul{\bbeta \balpha}} = R(\A)_{\ul{\bbeta \balpha}} = 0
\end{gather}
\item {Dimension 3/2}
\begin{gather}
R(\D)_\beta{}^j{}_{\,\alpha \dalpha} = +\frac{i}{2} \eps_{\beta \alpha} \bar\nabla^{\dphi j}
     \bar W_{\dphi \dalpha} \\
R(\D)_\dbeta{}^j{}_{\,\alpha \dalpha} = -\frac{i}{2} \eps_{\dbeta \dalpha} \nabla^{\phi j} W_{\phi \alpha} \\
R(\A)_\beta{}^j{}_{\,\alpha \dalpha} = -\frac{1}{4} \eps_{\beta \alpha} \bar\nabla^{\dphi j}
     \bar W_{\dphi \dalpha} \\
R(\A)_\dbeta{}^j{}_{\,\alpha \dalpha} = -\frac{1}{4} \eps_{\dbeta \dalpha} \nabla^{\phi j} W_{\phi \alpha}
\end{gather}
\item{Dimension 2}
\begin{align}
R(\D)_{\beta \dbeta\,\alpha \dalpha} &=
    \frac{1}{8} \eps_{\dbeta \dalpha} \left(\nabla_\beta{}^\phi W_{\phi \alpha}
          + \nabla_\alpha{}^\phi W_{\phi \beta}\right)
     - \frac{1}{8} \eps_{\beta \alpha} \left(\bar\nabla_{\dbeta \dphi} \bar W^\dphi{}_\dalpha
          + \bar\nabla_{\dalpha \dphi} \bar W^\dphi{}_\dbeta\right) \\
R(\A)_{\beta \dbeta\,\alpha \dalpha} &=
    - \frac{i}{16} \eps_{\dbeta \dalpha} \left(\nabla_\beta{}^\phi W_{\phi \alpha}
          + \nabla_\alpha{}^\phi W_{\phi \beta}\right)
     - \frac{i}{16} \eps_{\beta \alpha} \left(\bar\nabla_{\dbeta \dphi} \bar W^\dphi{}_\dalpha
          + \bar\nabla_{\dalpha \dphi} \bar W^\dphi{}_\dbeta\right)
\end{align}
\end{itemize}

\subsection{Isospin curvature}
The isospin curvature two-form is
\begin{align}\label{eq_RI2form}
R(I)_{ij} = d\Phi_{ij} - \Phi^k{}_{(i} \wedge \Phi_{j) k}
     + 4 E^\beta{}_{(i} \wedge F_{\beta j)} - 4 E_{\dbeta (i} \wedge F^\dbeta{}_{j)}
\end{align}
We group the results by dimension.
\begin{itemize}
\item {Dimension 1}
\begin{gather}
R(I)_{\ul{\bbeta \balpha} \,ij} = 0
\end{gather}
\item {Dimension 3/2}
\begin{gather}
R(I)_\beta{}^k{}_{\alpha \dalpha\,}{}_{ij} =
     - \frac{i}{2} \eps_{\beta \alpha} \delta^k_j \bar\nabla^\dphi{}_i \bar W_{\dphi \dalpha}
     - \frac{i}{2} \eps_{\beta \alpha} \delta^k_i \bar\nabla^\dphi{}_j \bar W_{\dphi \dalpha} \\
R(I)_\dbeta{}^k{}_{\alpha \dalpha\,}{}_{ij} = 
     + \frac{i}{2} \eps_{\dbeta \dalpha} \delta^k_j \nabla^\phi{}_i W_{\phi \alpha}
     + \frac{i}{2} \eps_{\dbeta \dalpha} \delta^k_i \nabla^\phi{}_j W_{\phi \alpha}
\end{gather}
\item{Dimension 2}
\begin{align}
R(I)_{\beta \dbeta\,\alpha \dalpha\,\,ij} =
     - \frac{1}{4} \eps_{\dbeta \dalpha} \nabla_{ij} W_{\beta \alpha}
     + \frac{1}{4} \eps_{\beta \alpha} \bar\nabla_{ij} \bar W_{\dbeta \dalpha}
\end{align}
\end{itemize}

\subsection{Special conformal curvature}
The special conformal curvatures are
\begin{align}
R(K)^a &= dF^a - F^b \wedge \Omega_b{}^a - F^a \wedge B
     - 2i F^{\alpha j} \wedge F^\dalpha{}_j \,\sigma^a_{\alpha \dalpha} \\
R(K)^{\alpha i} &= dF^{\alpha i}
     - \frac{1}{2} F^{\alpha i} \wedge B
     + i F^{\alpha i} \wedge A + F^{\beta i} \wedge \Omega_\beta{}^\alpha
     - F^{\alpha j} \wedge \Phi_j{}^i
     + i E_{\dalpha}{}^i \wedge F^b \,\bsigma_b^{\dalpha \alpha} \\
R(K)_{\dalpha i} &= dF_{\dalpha i}
     - \frac{1}{2} F_{\dalpha i} \wedge B
     - i F_{\dalpha i} \wedge A + F_{\dbeta i} \wedge \Omega^\dbeta{}_\dalpha
     + F_{\dalpha j} \wedge \Phi^j{}_i
     + i E^{\alpha}{}_i \wedge F_b \,\sigma^b_{\alpha \dalpha}
\end{align}
The components $R(K)_{\ul \bgamma \ul \bbeta}{}^A$ are given by
\begin{gather}
R(K)_{\bgamma \bbeta}{}^{\alpha j} =
     R(K)^{\dbgamma \dbbeta}{}_{\dalpha j} = R(K)_{\bgamma}{}^{\dbbeta}{}^A = 0 \\
R(K)_{\bgamma \bbeta}{}_{\dalpha}{}^j =
     \frac{1}{2} C_{\bgamma \bbeta} \bar\nabla_\dphi{}^j \bar W^\dphi{}_\dalpha, \quad
R(K)^{\dbgamma \dbbeta}{}_{\alpha}{}^j =
     -\frac{1}{2} C^{\dbgamma \dbbeta} \nabla^{\phi j} W_{\phi \alpha} \\
R(K)_{\bgamma \bbeta \,\,\alpha \dalpha} =
     - C_{\bgamma \bbeta} \nabla_{\alpha \dphi} \bar W^\dphi{}_\dalpha, \quad
R(K)^{\dbgamma \dbbeta}{}_{\,\,\alpha \dalpha} =
     + C^{\dbgamma \dbbeta} \nabla_{\dalpha}{}^\phi W_{\phi \alpha}
\end{gather}
The components
$R(K)_{\ul \bgamma b}{}^A = -\frac{1}{2} \bsigma_b^{\dbeta \beta} R(K)_{\ul \bgamma \,\beta \dbeta}{}^A$
are given by
\begin{gather}
R(K)_\gamma{}^j{}_{\,\beta \dbeta\,}{}_\alpha{}^k =
     \frac{1}{2} \eps_{\gamma\beta} \eps^{jk} \nabla_{\alpha \dphi} \bar W^\dphi{}_\dbeta, \quad
R(K)_\dgamma{}^j{}_{\,\beta \dbeta\,}{}_\dalpha{}^k =
     -\frac{1}{2} \eps_{\dgamma\dbeta} \eps^{jk} \nabla_{\dalpha}{}^\phi W_{\phi \beta} \\
R(K)_\gamma{}^j{}_{\,\beta \dbeta\,}{}_\dalpha{}^k =
     -\frac{i}{4} \eps_{\gamma \beta} \bar\nabla_\dbeta^j \bar\nabla_\dphi^k \bar W^\dphi{}_\dalpha, \quad
R(K)_\dgamma{}^j{}_{\,\beta \dbeta\,}{}_\alpha{}^k =
     -\frac{i}{4} \eps_{\dgamma \dbeta} \nabla_\beta^j \nabla^{\phi k} W_{\phi \alpha} \\
R(K)_\gamma{}^j{}_{\,\beta \dbeta\,\,\alpha \dalpha}
     = \frac{i}{2} \eps_{\gamma \beta} \bar\nabla^j_\dbeta \nabla_{\alpha \dphi} \bar W^\dphi{}_\dalpha, \quad
R(K)_\dgamma{}^j{}_{\,\beta \dbeta\,\,\alpha \dalpha}
     = \frac{i}{2} \eps_{\dgamma \dbeta} \nabla^j_\beta \nabla_{\dalpha}{}^\phi W_{\phi \alpha}
\end{gather}
The components $R(K)_{cb}{}^A = \frac{1}{4} \bsigma_c^{\dgamma \gamma} \bsigma_b^{\dbeta \beta}
     R(K)_{\gamma \dgamma\,\, \beta \dbeta}{}^A$ are given by
\begin{align}\label{eq_ScurvApp}
R(K)_{\gamma \dgamma \,\beta \dbeta \,\alpha}{}^i &=
     \frac{1}{48} \eps_{\dgamma \dbeta}
          \left(\nabla^{ij} \nabla_{\gamma j} W_{\beta \alpha}
               + \nabla^{ij} \nabla_{\beta j} W_{\gamma \alpha}
          \right)
     \eol & \quad
     + \eps_{\gamma \beta}
          \left(
               \frac{i}{8} \nabla_\alpha{}^\dphi \bar\nabla_\dgamma{}^j \bar W_{\dbeta \dphi}
               + \frac{i}{8} \nabla_\alpha{}^\dphi \bar\nabla_\dbeta{}^j \bar W_{\dgamma \dphi}
               + \frac{1}{4} \bar W_{\dgamma \dbeta}  \nabla^{\phi j} W_{\phi \alpha}
          \right) \\  \label{eq_SbcurvApp}
R(K)_{\gamma \dgamma \,\beta \dbeta \,\dalpha\, i} &=
     \frac{1}{48} \eps_{\gamma \beta}
          \left(\bar \nabla_{ij} \bar \nabla_{\dgamma}{}^j \bar W_{\dbeta \dalpha}
               + \bar \nabla_{ij} \bar \nabla_{\dbeta}{}^j \bar W_{\dgamma \dalpha}
          \right)
     \eol & \quad
     + \eps_{\dgamma \dbeta}
          \left(
               - \frac{i}{8} \nabla_\dalpha{}^\phi \nabla_{\gamma j} W_{\beta \phi}
               - \frac{i}{8} \nabla_\dalpha{}^\phi \nabla_{\beta j} W_{\gamma \phi}
               + \frac{1}{4} W_{\gamma \beta}  \bar \nabla^{\dphi}{}_j \bar W_{\dphi \dalpha}
          \right) \\ \label{eq_KcurvApp}
R(K)_{\gamma \dgamma\, \beta \dbeta\, \alpha \dalpha}
     &=  -\frac{1}{8} \eps_{\dgamma \dbeta} \nabla_\dalpha{}^\phi \nabla_{\gamma \beta} W_{\phi \dalpha}
          - \frac{1}{8} \eps_{\gamma \beta} \nabla_\alpha{}^\dphi \bar\nabla_{\dgamma \dbeta} \bar W_{\dphi \dalpha}
          \eol & \quad
          + \frac{1}{4} \eps_{\dgamma \dbeta} \nabla_\gamma{}^\dphi (\bar W_{\dphi \dalpha} W_{\beta \alpha})
          + \frac{1}{4} \eps_{\dgamma \dbeta} \nabla_\beta{}^\dphi (\bar W_{\dphi \dalpha} W_{\gamma \alpha})
          \eol & \quad
          + \frac{1}{4} \eps_{\gamma \beta} \nabla_\dgamma{}^\phi (W_{\phi \alpha} \bar W_{\dbeta \dalpha})
          + \frac{1}{4} \eps_{\gamma \beta} \nabla_\dbeta{}^\phi (W_{\phi \alpha} \bar W_{\dgamma \dalpha})
          \eol & \quad
          + \frac{i}{8} \eps_{\dgamma \dbeta} (\bar\nabla^\dphi{}_j \bar W_{\dphi \dalpha})
               (\nabla_\gamma{}^j W_{\beta \alpha} + \nabla_\beta{}^j W_{\gamma \alpha})
          \eol & \quad
          + \frac{i}{8} \eps_{\gamma \beta} (\nabla^{\phi j} W_{\phi \alpha})
               (\bar \nabla_{\dgamma j} \bar W_{\dbeta \dalpha}
                    + \bar \nabla_{\dbeta j} \bar W_{\dgamma \dalpha})
\end{align}
In the expression for $R(K)_{cba}$ we have reordered the derivatives given
in the original expression \eqref{eq_Kcurv}; where we originally had
$\nabla_{\gamma \beta} \nabla_\dalpha{}^\phi W_{\phi \dalpha}$, we have applied
the commutation relations to give
$\nabla_\dalpha{}^\phi \nabla_{\gamma \beta} W_{\phi \dalpha}$ (and similarly
for the complex conjugate). This simplifies the analysis since the bosonic derivative
is now furthest to the left.

\end{document}